\documentclass[twocolumn,secnumarabic,amssymb, nofootinbib, nobibnotes, aps, prd, reprint,superscriptaddress]{revtex4-1}

\usepackage{bm}           
\usepackage[cmex10]{amsmath}
\usepackage{amssymb}
\usepackage{mathrsfs}
\usepackage{mathtools}
\usepackage{feynmp-auto}
\usepackage[british]{babel}
\usepackage[titletoc,toc,title]{appendix}
\usepackage{scrextend}
\usepackage[usenames, dvipsnames]{color}

\begin{document}
\title{Effective action approach to wave propagation in scalar QED plasmas}
\author         {Yuan Shi}
\email          {Corresponding author: yshi@pppl.gov}
\affiliation    {Department of Astrophysical Sciences, Princeton University, Princeton, NJ 08544 USA}
\affiliation    {Princeton Plasma Physics Laboratory, Princeton University, Princeton, NJ 08543 USA}

\author         {Nathaniel J. Fisch}
\affiliation    {Department of Astrophysical Sciences, Princeton University, Princeton, NJ 08544 USA}
\affiliation    {Princeton Plasma Physics Laboratory, Princeton University, Princeton, NJ 08543 USA}

\author         {Hong Qin}
\affiliation    {Department of Astrophysical Sciences, Princeton University, Princeton, NJ 08544 USA}
\affiliation    {Princeton Plasma Physics Laboratory, Princeton University, Princeton, NJ 08543 USA}
\affiliation    {School of Nuclear Science and Technology, University of Science and Technology of China, Hefei, Anhui 230026, China}
\date           {\today}

\begin{abstract}
A relativistic quantum field theory with nontrivial background fields is developed and applied to study waves in plasmas. The effective action of the electromagnetic 4-potential is calculated ab initio from the standard action of scalar QED using path integrals. The resultant effective action is gauge invariant and contains nonlocal interactions, from which gauge bosons acquire masses without breaking the local gauge symmetry. To demonstrate how the general theory can be applied, we study a cold unmagnetized plasma and a cold uniformly magnetized plasma. Using these two examples, we show that all linear waves well-known in classical plasma physics can be recovered from relativistic quantum results when taking the classical limit. In the opposite limit, classical wave dispersion relations are modified substantially. In unmagnetized plasmas, longitudinal waves propagate with nonzero group velocities even when plasmas are cold. In magnetized plasmas, anharmonically spaced Bernstein waves persist even when plasmas are cold. These waves account for cyclotron absorption features observed in spectra of X-ray pulsars. Moreover, cutoff frequencies of the two non-degenerate electromagnetic waves are redshifted by different amounts. These corrections need to be taken into account in order to correctly interpret diagnostic results in laser plasma experiments.  
\end{abstract}

\maketitle
\setlength{\parskip}{0pt}
\section{Introduction}
Classical plasma theories, which describe plasmas as collections of charged classical particles moving in self-consistent electromagnetic (EM) fields, become deficient when relativistic and quantum effects are important. The conditions at which relativistic quantum effects are important may be determined by comparing three energy scales: the energy scales of the plasmas, the energy scales of the EM fields, and the rest energy of charged particles. The energy scales of plasmas are the thermal energy $k_{\text{B}}T$, the Fermi energy $\epsilon_\text{F}$, and the plasmon energy $\hbar\omega_p$. The energy scales of wave fields are the photon energy $\hbar\omega$ and the ponderomotive energy $U_p$. The energy scales of static electric and magnetic fields are $\epsilon_\text{E}=\sqrt{eEc\hbar}$ and $\epsilon_{\text{B}}=\sqrt{eBc^2\hbar}$, respectively. Relativistic effects are important when the energy scales of either the plasmas or the EM fields are comparable to the rest energy of charged particles. Quantum effects are important when the thermal energy is low compared to other energy scales. 

An example where relativistic quantum effects are important is the magnetosphere of an X-ray pulsar \cite{Meszaros92}. The typical magnetic fields of X-ray pulsars are on the order $B\sim10^{12}$ G. The corresponding magnetic energy $\epsilon_\text{B}\sim100$ KeV is comparable to the rest energy of electrons $m_ec^2\approx511$ KeV, indicating that relativistic effects are important. Moreover, the effective temperature of X-ray pulsars $k_{\text{B}}T\sim10$ KeV is colder than $\epsilon_\text{B}$, indicating that quantum effects are also important. That relativistic quantum effects are important, an inference made by comparing energy scales, is strongly supported by anharmonic cyclotron absorption features observed in spectra of X-ray pulsars \cite{Heindl00,Santangelo99,Tsygankov06,Tsygankov07,Pottschmidt05,Makishima90}. Since classical plasma theories cannot explain these spectral features, the presence of high order harmonics is attributed to inelastic scatterings of photons with electrons that occupy quantized Landau levels \cite{Harding06}, and the anharmonicity is attributed to viewing geometry as well as relativistic effects \cite{Schonherr07,Nishimura13}. Despite numerous efforts, many features of cyclotron absorption lines remain to be explained \cite{Meszaros85,Freeman99,Bignami03}. The locations and shapes of these lines contain important information such as the magnetic field and plasma profiles of magnetospheres of X-ray pulsars. This information cannot be extracted, unless wave dispersion relations, which enter the radiation transport equations \cite{Meszaros92} that serve as the forward model in the retrieval problem \cite{Rodgers00}, are obtained for strongly magnetized plasmas. In this paper, we will obtain, for the first time, explicit expressions of wave dispersion relations in strongly magnetized plasmas.

Another example where relativistic quantum effects are important is a plasma produced by ultra intense lasers interacting with a solid target. During the interaction, a quasistatic magnetic field on gigagauss scale can be produced \cite{Stamper91,Korneev15}. The corresponding magnetic energy $\epsilon_\text{B}\sim 1$ KeV is comparable to the electron temperature of the plasma, indicating that quantum effects are important. Relativistic effects also turn out to be important when optical lasers are used to diagnose the plasma. This is because the frequencies of optical photons are close to wave cutoffs, if the plasma has density $n_0\sim 10^{21}\hspace{3pt}\text{cm}^{-3}$, corresponding to $\hbar\omega_p\sim 1$ eV. Due to singularities near cutoffs, small modifications of the cutoff frequencies can have large effects. Such effects have been revealed in a number of experiments \cite{Tatarakis02,Wagner04}. In these experiments, it is found that the magnetic field, determined from classical formulas, is larger when the same plasma is diagnosed by lasers with higher frequencies. This peculiar dependence of the inferred magnetic field strength on the frequencies of diagnostic lasers indicates that systematic errors exist in classical formulas. These systematic errors can be removed only by carefully calculating how waves propagate in strongly magnetized plasmas. In this paper, we will describe, for the first time, how Faraday rotation is modified by relativistic quantum effects.

To describe wave propagation in strongly magnetized plasmas, as well as in other plasmas where relativistic quantum effects are important, we develop a general theory that can be applied to various situations. The general theory, which is based on the concept of effective action in quantum field theory, describes wave propagation through possibly nonthermal scalar QED plasmas in background EM fields. The wave effective action has a clear physical meaning. When waves propagate through plasmas, they interact with charged particles, whose dynamics are affected by the presence of the background fields as well as the wave fields. After all the interactions related to charged particles are summed up, what remains is the effective action of waves. This clear physical picture of the wave effective action can be translated into rigorous mathematical procedures. To derive the effective action, we start from the standard action of scalar QED, self-consistently factor out the background fields from the fluctuating fields, and then integrate out the charged particle fields in the path integral perturbatively. We shall see that at the 1-loop level, the effective action is related to the linear response of the system to small amplitude waves. 

Having developed the general theory, we demonstrate how it can be applied using two examples. The first example is a uniform, unmagnetized, boson plasma. Such a system has been studied by a number of authors using other methods \cite{Hines78, Kowalenko85}. Results of our general theory agree with these authors' in this special case. The second example is a uniform, magnetized, boson plasma. Such a system has been studied by Witte and coauthors \cite{Witte87, Witte88, Witte90}, who can only describe wave propagation parallel to the background magnetic field. Our results agree in this special case. Moreover, our transparent formalism also enables us to describe nonparallel wave propagation, which was obscure in previous studies. Although many theories and models have been developed in the literature to describe waves in relativistic quantum plasmas, such as plasma response theories \cite{Melrose07, Melrose12}, finite temperature theories \cite{Kapusta06,Landsman87,Inagaki05}, quantum hydrodynamic models \cite{Haas11, Shukla10}, and models that are based on the historical Heisenberg-Euler effective Lagrangian \cite{Heisenberg36,Bialynicka70,Marklund06,Piazza07,Lundin09}, this is the first theory capable of demonstrating its correctness by showing that all linear wave modes well-known in classical plasma theories can be recovered when taking the classical limit\footnote{The classical regime is $mc^2\gg pc\gg \epsilon$, where $m$ is the rest mass, $p$ is the momentum, and $\epsilon$ is the quantized energy of a particle.} in relativistic quantum results.

Beyond the immediate goal of establishing a formalism that describes wave propagation in relativistic quantum plasmas, the goal of this paper is to demonstrate that quantum field theory, in whose language the standard model of particle physics is written, and in whose language many phenomena in condensed matter physics are explained, is also an effective language for plasma physics. Since particle physics describes a few particles with high energy, condensed matter physics describes many particles with low energy, and plasma physics describes intermediate number of particles with intermediate energy, it should not be surprising that a language that is effective for both extremes is also effective in the intermediate regime. In this paper, we study scalar QED, which describes spin-0 bosons, instead of spinor QED, which describes spin-1/2 fermions, so that our main ideas can be illustrated without complications due to spins. Although plasmas are typically made of spinor particles instead of scalar particles, scalar QED can be a good model for many condensed matter systems like superconductors and superfluids \cite{Landau65}. Moreover, our scalar QED plasma theory does recover waves in classical plasma theories, which typically take no account of spins at all. Extension of our formalism to spinor QED plasmas is straightforward and will be reported separately. 

The rest of this paper is organized as follows. In Sec. \ref{sec:action}, the general theory of wave propagation in scalar QED plasmas is presented. In Sec. \ref{sec:unmagnetized}, we evaluate the wave effective action in unmagnetized plasmas and obtain linear wave modes in a cold plasma. In Sec. \ref{sec:magnetized}, an explicit expression of the wave effective action in a cold and uniformly magnetized plasma is obtained, followed by analysis of linear wave modes. Conclusions and discussion are made in Sec. \ref{sec:conclusion}. Some important mathematical details are given in the Appendices.

\section{\label{sec:action}General Theory}
\subsection{Quantum field theory with background fields}
Our starting point is the standard action of scalar QED, in which a complex scalar field is coupled to the gauge field through minimum coupling. In natural unit $\epsilon_0=\hbar=c=1$ and Minkowski metric $g^{\mu\nu}=\text{diag}(1,-1,-1,-1)$, the standard scalar QED action is
\begin{eqnarray}\label{eq:action}
\nonumber
S=\int d^4x&\bigg(&(D_{\mu}\phi)^{*}(D^{\mu}\phi)-m^2\phi^{*}\phi\\
&-&\frac{\lambda}{2}(\phi^{*}\phi)^2-\frac{1}{4}F_{\mu\nu}F^{\mu\nu}\bigg).
\end{eqnarray}
The complex scalar field $\phi$, whose complex conjugate is denoted as $\phi^*$, describes charged spin-0 bosons with mass $m$ and charge\footnote{In natural unit, the electron charge $e=-|e|$ is a dimensionless small number related to the fine-structure constant $\alpha=e^2/4\pi\approx1/137$. We denote $e_s$ the charge of particle states of species $s$, so the charge of anti-particle states is $-e_s$.} $e$. The real-valued 1-form $A=A_{\mu}dx^{\mu}$ is the gauge field that defines the gauge covariant derivative $D_{\mu}=\partial_{\mu}-ieA_{\mu}$. The covariant derivative has curvature 2-form $F_{\mu\nu}=\partial_{\mu}A_{\nu}-\partial_{\nu}A_{\mu}$, known as the field strength tensor. The $\phi^4$ interaction is necessary for the theory to be renormalizable, and the self-coupling constant $\lambda$ is in general nonzero. But to 1-loop level, $\lambda$ does not contribute to the propagation of gauge bosons. For simplicity, we set the renormalized value of $\lambda$ to zero from now on. It is well-known that the Lagrangian density of action (\ref{eq:action}) is invariant under local U(1)-gauge transformation
\begin{equation}\label{Gauge}
\phi\rightarrow\phi e^{ie\chi},\quad A_{\mu}\rightarrow A_{\mu}+\partial_{\mu}\chi,
\end{equation}
where $\chi$ is any function. The gauge invariant symmetry current is
\begin{equation}\label{current}
J^{\mu}=\frac{e}{i}[\phi^{*}(D^{\mu}\phi)-\phi(D^{\mu}\phi)^{*}].
\end{equation}
Contributions by other charged species are additive to the current as well as to the Lagrangian. For conciseness, we will not write summations over charged species explicitly. 

\begin{figure}
	\renewcommand{\figurename}{FIG.}
	\includegraphics[angle=0,width=8.5cm]{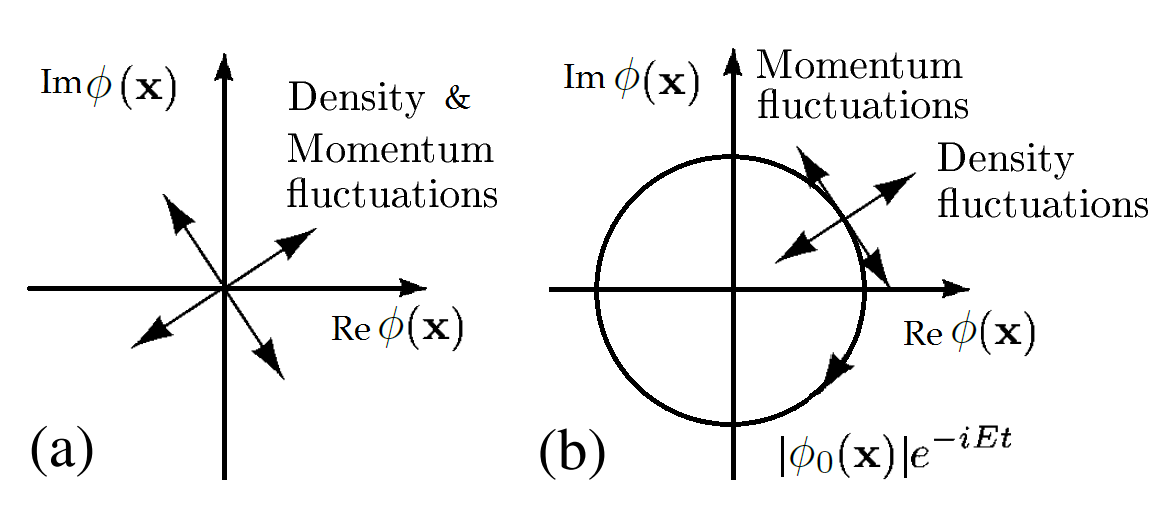}
	\caption{Comparison between field theory in the vacuum and field theory with dynamical backgrounds. When a field $\phi(\mathbf{x})$ fluctuates near the vacuum (a), energy and momentum fluctuations are always accompanied by density fluctuations. On the other hand, when the field fluctuates near some non-vacuum background (b), energy and momentum fluctuations can be orthogonal to density fluctuations.}
	\label{fig:background}
\end{figure}

To describe plasmas, which are constituted of charged particles and their self-consistent EM fields, let us first understand the roles of background fields in quantum field theory. In the usual quantum field theory, fields fluctuate near their vacuum expectation values. The creation and annihilation of energy momentum quanta are always accompanied by the creation and annihilation of particles. However, when fields fluctuate near some non-vacuum backgrounds, these two types of fluctuations become orthogonal. Energy momentum quanta can be created and annihilated without changing the number of particles in a system, and the number of particles can be altered without changing the total energy and momentum of the system. This idea is illustrated in Fig. \ref{fig:background}. Mathematically, the fields $\phi$ and $A$ can be decomposed into classical backgrounds and quantum fluctuations
\begin{equation}\label{Expand}
\phi=\phi_{0}+\varphi,\quad A_{\mu}= \bar{A}_{\mu}+\mathcal{A}_{\mu}.  
\end{equation}
Vacuum is the trivial case when the background fields $\phi_0$ and $\bar{A}$ are zero. When the background fields are nontrivial, the self-consistency of the classical backgrounds is given by the Euler$-–$Lagrange equations 
\begin{eqnarray}\label{EOM}
\nonumber
(\bar{D}_{\mu}\bar{D}^{\mu}+m^2)\phi_{0}&=&0,\\
\partial_{\mu}\bar{F}^{\mu\nu}=\bar{J}_{0}^{\nu}.
\end{eqnarray}
In the above equations, $\bar{D}_{\mu}=\partial_{\mu}-ie\bar{A}_{\mu}$ is the background gauge covariant derivative, $\bar{F}_{\mu\nu}=\partial_{\mu}\bar{A}_{\nu} -\partial_{\nu}\bar{A}_{\mu}$ is the background field strength tensor, and $\bar{J}_{0}^{\nu}=\sum_s\bar{J}_{s0}^{\nu}$ is the total background current, summed over all charged species. It is clear that the above equations are invariant under the background U(1)-gauge transformation on $\phi_{0}$ and $\bar{A}$. The classical equations of motion (\ref{EOM}) describe bound states as well as unbound states. When the potential energy is larger than the kinetic energy, as is the case in condensed matter systems, particles are bound by the potential created by other particles. In this case, the wave functions $\phi_0$ and $\bar{A}$ are localized and interactions between particles can be strong. On the other hand, when the kinetic energy is larger than the potential energy, as is the case in plasmas, particles are unbound. In this case, the motion of one particle is weakly correlated with the motion of other particles, except during collisions.

While the background EM field $\bar{F}_{\mu\nu}$ is conceptually simple, the background charged particle field $\phi_0$ needs some clarifications. When the plasma background is constituted of N bosons, the classical background field $\phi_{0}(x)$ is formally related to the properly symmetrized N-body wave function $\Phi_{0}(x_1,x_2,\dots,x_N)$ by
\begin{equation}
\phi_0(x)=\int\sqrt{V}\Phi_0(x,x_2,\dots,x_N).
\end{equation}
Here $V=d^4x_2\wedge\dots\wedge d^4x_N$ is the volume form of the $4(N-1)$ dimensional subspace space of the N-boson configuration space.  The half-form $\sqrt{V}$ is commonly seen in geometric quantization \cite{Bates97}. It is easy to check that the N-body wave function has mass dimension $[\Phi_0(x_1,\dots,x_N)]=M^{2N-1}$ and the field $\phi_0(x)$ has mass dimension $[\phi_0(x)]=M$ as expected. When a pair $\phi_{0}(x)\phi^{*}_{0}(x)$ appears in an expression, two half-forms combine into the volume form, and the integration can then be carried out. For example, the current $\bar{J}_{0}^{\mu}$ of an $N$-body wave function $\Phi_{0}$ can be written explicit as
\begin{equation}\label{MBcurrent}
\bar{J}_{0}^{\mu}(x)=\frac{e}{i}\!\int\!V[\Phi_{0}^{*}(x,x_2...)\bar{D}^{\mu}(x)\Phi_{0}(x,x_2...)-c.c.].
\end{equation}
For conciseness, whenever the pair $\phi_{0}(x)\phi^{*}_{0}(x)$ appears in an expression, integration over all other coordinates of $x_2,\dots$ of the many-body wave function $\Phi_0(x,x_2,\dots)$ will be implied.

Having clarified the roles of background fields, let us study their effects in quantum field theory. Separate classical backgrounds from quantum fluctuations using decomposition (\ref{Expand}), integrate by part using 
\begin{equation}\label{eq:IBT}
\int dx h^{*}\bar{D}_{\mu}f=-\int dx(\bar{D}_{\mu}h)^{*}f,
\end{equation} 
and cancel terms using the background self-consistency conditions (\ref{EOM}), we can write action (\ref{eq:action}) as
\begin{eqnarray}\label{ReducedAction}
\nonumber
S=\int d^4x&\bigg(&(\bar{D}_{\mu}\varphi)^{*}(\bar{D}^{\mu}\varphi)-m^2\varphi^{*}\varphi\\
\nonumber
&-&\frac{1}{4}(\bar{F}_{\mu\nu}\bar{F}^{\mu\nu}+\mathcal{F}_{\mu\nu}\mathcal{F}^{\mu\nu})\\
&-&(\bar{\jmath}^{\mu}+\bar{\eta}^{\mu})\mathcal{A}_{\mu}+e^2\phi\phi^{*}\mathcal{A}_{\mu}\mathcal{A}^{\mu}\bigg).
\end{eqnarray}
Here $\mathcal{F}_{\mu\nu}=\partial_{\mu}\mathcal{A}_{\nu}-\partial_{\nu}\mathcal{A}_{\mu}$ is the field strength tensor for the fluctuating field $\mathcal{A}$,
\begin{eqnarray}
\bar{\jmath}^{\mu}&:=&\frac{e}{i}(\varphi^{*}\bar{D}^{\mu}\varphi-c.c.),\\
\bar{\eta}^{\mu}&:=&\frac{e}{i}(\phi_{0}^{*}\bar{D}^{\mu}\varphi+\varphi^{*}\bar{D}^{\mu}\phi_{0}-c.c.),
\end{eqnarray}
are currents due to vacuum excitation and background plasma excitation, respectively.

The background-reduced action (\ref{ReducedAction}) is associated with a Lagrangian density $\mathcal{L}$ that depends on the fluctuating fields $\varphi$ and $\mathcal{A}$. When waves propagate through background plasmas, the background field $\bar{F}_{\mu\nu}$, which is usually generated by some slowly-varying and large-scale external charge current distributions, can be regarded as fixed. The background charged particle field $\phi_{0}$, which is self-consistent with $\bar{F}_{\mu\nu}$, can also be regarded as fixed. In this way, all interactions between waves and charged particles are taken into account by the fluctuating fields. Up to the constant term $\frac{1}{4}\bar{F}_{\mu\nu}\bar{F}^{\mu\nu}$, the Lagrangian density of the fluctuating fields $\mathcal{A}$ and $\varphi$ is
\begin{eqnarray}\label{ReducedLagrangian}
\nonumber
\mathcal{L}\enspace=&&\mathcal{L}_{\varphi}+\mathcal{L}_{\mathcal{A}}+\mathcal{L}_{I}\\
\nonumber
=&&(\bar{D}_{\mu}\varphi)^{*}(\bar{D}^{\mu}\varphi)-m^2\varphi^{*}\varphi\\
&-&\frac{1}{4}\mathcal{F}_{\mu\nu}\mathcal{F}^{\mu\nu}+e^2\phi_{0}^{*}\phi_{0}\mathcal{A}_{\mu}\mathcal{A}^{\mu}\\
\nonumber
&-&(\bar{\jmath}^{\mu}+\bar{\eta}^{\mu})\mathcal{A}_{\mu}+e^2(\phi_{0}\varphi^{*}+\phi_{0}^{*}\varphi +\varphi^{*}\varphi)\mathcal{A}_{\mu}\mathcal{A}^{\mu}.
\end{eqnarray}
Here $\mathcal{L}_{\varphi}$, $\mathcal{L}_{\mathcal{A}}$, and $\mathcal{L}_{I}$ correspond to terms on the second, the third, and the fourth line, respectively. $\mathcal{L}_{\varphi}$ is the Lagrangian density of the free $\varphi$ field. It should be clarified that $\varphi$ is not free in the sense that its dynamics is influenced by the background field $\bar{A}$, as is manifested by the background gauge covariant derivative $\bar{D}$ acting on $\varphi$. But $\varphi$ is nevertheless free in the sense that it neither interacts with $\mathcal{A}$ nor couples to itself. Similarly, $\mathcal{L}_{\mathcal{A}}$ is the Lagrangian density of the free $\mathcal{A}$ field. Notice that the background field $\phi_0$ endows the gauge field $\mathcal{A}$ with a mass term that can have spatial and temporal dependences. Finally, the interaction Lagrangian $\mathcal{L}_{I}$ contains interactions between $\varphi$ and $\mathcal{A}$. Some interactions only involve the fluctuating fields $\varphi$ and $\mathcal{A}$ with constant couplings. These interactions happen in plasmas as well as in the vacuum. Other interactions involve the background fields $\phi_0$ and $\bar{A}$ in the coupling. These interactions do not happen unless nontrivial background fields are present.

The Lagrangian density (\ref{ReducedLagrangian}) has a number of gauge symmetries. It is obvious that the Lagrangian is invariant under background local U(1)-gauge transformation
\begin{equation}\label{BackgroundGauge}
\phi_{0}\rightarrow\phi_{0} e^{ie\chi},\quad \varphi\rightarrow\varphi e^{ie\chi} ,\quad \bar{A}_{\mu}\rightarrow \bar{A}_{\mu}+\partial_{\mu}\chi.
\end{equation}
It is less obvious but equally true that once the background fields $\phi_{0}$ and $\bar{A}$ are fixed the Lagrangian is invariant under the following transformation
\begin{equation}\label{Symmetry}
\mathcal{A}_{\mu}\rightarrow \mathcal{A}_{\mu}+\partial_{\mu}\chi ,\quad \varphi\rightarrow\varphi e^{ie\chi}+\phi_{0}(e^{ie\chi}-1).
\end{equation}
This can be understood intuitively from Fig. \ref{fig:background} as follows. The local U(1)-gauge transformation (\ref{Gauge}) is a shift in $A$ and a phase rotation in $\phi$. When $\bar{A}$ is fixed the shift is completely absorbed into $\mathcal{A}$. When $\phi_{0}$ is fixed $\varphi$ has to transform by Eq. (\ref{Symmetry}) in order to preserve the norm of $\phi$. The symmetry current is
\begin{equation}
\mathcal{J}^{\mu}=\bar{\jmath}^{\mu}+\bar{\eta}^{\mu}-2e^2\phi\phi^{*}\mathcal{A}^{\mu}.
\end{equation}
This gauge invariant current $\mathcal{J}$ contains contributions from excitations of the background fields as well as excitations of the vacuum.

\subsection{Effective action of gauge bosons}
So far, no approximation has been made, and the Lagrangian density (\ref{ReducedLagrangian}) is exact. The Lagrangian density describes the free $\varphi$ field, the free $\mathcal{A}$ field, and their interactions. When the gauge field $\mathcal{A}$ propagates, it interacts with charged particles and becomes dressed by these interactions. After summing up all these dressings, the effective action, which describes the propagation of the dressed $\mathcal{A}$ field, can be obtained. The summation of dressings can be rigorously implemented using the path integral\footnote{The path integral is one way of quantizing the field theory. It superimposes all possible field configurations weighted by the phase factor $e^{iS/\hbar}$.}, which can be evaluated perturbatively using the small dimensionless coupling constant $e$ as an expansion parameter. 

Formally, the exponentiated effective action $e^{i\Gamma\bm{[}\mathcal{A}\bm{]}}$ of the $\mathcal{A}$ field is the partially evaluated quantum partition function\footnote{The quantum partition function serves a similar role as the statistical partition function. In the statistical case, the average is weighted by the Boltzmann factor $e^{-H/k_\text{B}T}$, while in the quantum case, the average is weighted by the phase factor $e^{iS/\hbar}$.} when the $\varphi$ field is integrated out. To integrate out the $\varphi$ field, we will need to expand the action exponential $e^{iS}$, the quantum equivalence of the Boltzmann factor,  and it is convenient to group terms in the interaction $S_{I}=\int d^4x\mathcal{L}_{I}$ according to their powers in $e$, $\varphi$, and $\mathcal{A}$. Schematically, we can write
\begin{equation}
S_{I}=S_{e\varphi\mathcal{A}}+S_{e\varphi^2\mathcal{A}}+S_{e^2\varphi\mathcal{A}^2}+S_{e^2\varphi^2\mathcal{A}^2}.
\end{equation}
Denote the action of the free $\varphi$ field and the free $\mathcal{A}$ field by $S_\varphi$ and $S_\mathcal{A}$, respectively. Expand the action exponential to $e^2$ order and use properties of Gaussian integrals to eliminate terms that contain odd powers of $\varphi$ in the path integral, we obtain the exponentiated effective action
\begin{eqnarray}\label{EffectiveAction}
\nonumber
e^{i\Gamma\bm{[}\mathcal{A}\bm{]}}&:=&\frac{1}{Z_{\varphi}}\int[\mathscr{D}\varphi][\mathscr{D}\varphi^{*}]e^{i(S_{\varphi}+S_{\mathcal{A}}+S_{I})}\\
\nonumber
&=&e^{iS_{\mathcal{A}}}\frac{1}{Z_{\varphi}}\int[\mathscr{D}\varphi][\mathscr{D}\varphi^{*}]e^{iS_{\varphi}}\\
\nonumber
&&\hspace{30pt}\times\bigg[1+i\Big(S_{e\varphi^2\mathcal{A}}+S_{e^2\varphi^2\mathcal{A}^2}\Big)\\
&&\hspace{35pt}+\frac{i^2}{2}\Big(S_{e\varphi\mathcal{A}}^2+S_{e\varphi^2\mathcal{A}}^2\Big)+o(e^2)\bigg],
\end{eqnarray}
where $\int[\mathscr{D}\varphi][\mathscr{D}\varphi^{*}]$ denotes the path integral over fields $\varphi$ and $\varphi*$, and $Z_{\varphi}:=\int[\mathscr{D}\varphi][\mathscr{D}\varphi^{*}]e^{iS_{\varphi}}$ is the partition function of the free $\varphi$ field. The term $S_{e\varphi^2\mathcal{A}}$ is linear in $\mathcal{A}$. It serves as the source term that is responsible for the emission, absorption and scattering of gauge bosons. Since our focus is wave propagation, we will not be concerned with this term in this paper. The remaining terms in the expansion (\ref{EffectiveAction}) are quadratic in $\mathcal{A}$ and they are responsible for the propagation of the gauge field.

To express the effective action in a more illuminating form, we can write terms in the above expansion in terms of quantities that are familiar in quantum field theory. The first quantity is the propagator, or the Green's function, of the free $\varphi$ field
\begin{eqnarray}
\nonumber
G(x,x')&=&\langle\varphi(x)\varphi^*(x')\rangle_{\varphi}\\
&:=&\frac{1}{Z_{\varphi}}\int[\mathscr{D}\varphi][\mathscr{D}\varphi^{*}]e^{iS_{\varphi}}\varphi(x)\varphi^{*}(x').
\end{eqnarray}
The Green's function is a 2-point quantum correlation function. It is the probability amplitude of creating a charged particle at one location and then annihilating this particle at another location. The Green's function of the free  $\varphi$ field appears from $S_{e^2\varphi^2\mathcal{A}^2}$ after evaluating the path integral (\ref{EffectiveAction}). The second quantity is the gauge invariant polarization tensor $\Pi^{\mu\nu}(x,x')$. Use properties of Gaussian integrals to integrate out the $\mathcal{A}$ field\footnote{Integrating the $\mathcal{A}$ field requires gauge fixing. However, since the $\mathcal{A}$ field does not contribute to the polarization tensor to $e^2$ order, we do not need to be concerned with gauge fixing at this order.}, we can evaluate the exact polarization tensor as
\begin{eqnarray}\label{polarization}
\nonumber
\Pi^{\mu\nu}(x,x')&=&\langle\mathcal{J}^{\mu}(x)\mathcal{J}^{\nu}(x')\rangle\\
\nonumber
&:=&\frac{1}{Z}\int[\mathscr{D}\varphi][\mathscr{D}\varphi^{*}][\mathscr{D}\mathcal{A}]e^{iS}\mathcal{J}^{\mu}(x)\mathcal{J}^{\nu}(x')\\
\nonumber
&=&\frac{1}{Z_{\varphi}}\int[\mathscr{D}\varphi][\mathscr{D}\varphi^{*}]e^{iS_{\varphi}}(\bar{\eta}^{\mu}\bar{\eta}^{\nu}+\bar{\jmath}^{\mu}\bar{\jmath}^{\nu})+o(e^2)\\
&=&\Pi^{\mu\nu}_{2,\text{bk}}(x,x')+\Pi^{\mu\nu}_{2,\text{vac}}(x,x')+o(e^2).
\end{eqnarray}
where $Z=\int[\mathscr{D}\varphi][\mathscr{D}\varphi^{*}][\mathscr{D}\mathcal{A}]e^{iS}$ is the total partition function of Lagrangian density (\ref{ReducedLagrangian}). The polarization tensor is the quantum current-current correlation function. It is the probability amplitude that a current is excited at one location and then de-excited at another location. The two terms $\Pi^{\mu\nu}_{2,\text{bk}}(x,x')$ and $\Pi^{\mu\nu}_{2,\text{vac}}(x,x')$ are polarization of the background plasma and the polarization of the vacuum, respectively. They appear from $S_{e\varphi\mathcal{A}}^2$ and $S_{e\varphi^2\mathcal{A}}^2$ after evaluating the path integral (\ref{EffectiveAction}). The subscript ``2" indicates that they are approximated expressions to $e^2$ order in the perturbation series. 

In terms of these quantities, the effective action of gauge boson propagation to order $e^2$ can be written in the concise form
\begin{eqnarray}\label{1-loop}
\nonumber
\Gamma_2\bm{[}\mathcal{A}\bm{]}&=&\frac{1}{2}\int d^4x\Bigg(\mathcal{A}_{\mu}(x)(\partial^2g^{\mu\nu}-\partial^{\mu}\partial^{\nu})\mathcal{A}_{\nu}(x)\\
&&+\int d^4x'\mathcal{A}_{\mu}(x)\Sigma_2^{\mu\nu}(x,x')\mathcal{A}_{\nu}(x')\Bigg),
\end{eqnarray}
where $\Sigma_2^{\mu\nu}(x,x')$ is the response tensor to order $e^2$. The response tensor contains contributions from the background plasma as well as the vacuum
\begin{equation}
\Sigma_2^{\mu\nu}(x,x')=\Sigma_{2,\text{bk}}^{\mu\nu}(x,x')+\Sigma_{2,\text{vac}}^{\mu\nu}(x,x').
\end{equation}
The response due to the background plasma is constituted of the gauge boson mass term and the plasma polarization term
\begin{fmffile}{bk}
\begin{eqnarray}\label{bk}
\Sigma_{2,bk}^{\mu\nu}(x,x')
&=&\quad
\begin{gathered}
\begin{fmfgraph*}(40,15)
\fmfkeep{mass}
\fmfleft{i}
\fmfright{o}
\fmf{photon}{i,v}
\fmf{photon}{v,o}
\fmfdot{v}
\fmfv{label=$x$,label.angle=90,label.dist=6}{v}
\fmfv{label=$\mu$,label.dist=0.2}{i}
\fmfv{label=$\nu$,label.dist=0.2}{o}
\end{fmfgraph*}
\end{gathered}
\quad+\quad
\begin{gathered}
\begin{fmfgraph*}(40,15)
\fmfkeep{line}
\fmfleft{i}
\fmfright{o}
\fmf{plain}{v1,v2}
\fmfdot{v1,v2}
\fmfv{label=$x$,label.angle=90,label.dist=8}{v1}
\fmfv{label=$x'$,label.angle=90,label.dist=8}{v2}
\fmfv{label=$\mu$,label.dist=0.5}{i}
\fmfv{label=$\nu$,label.dist=0.5}{o}
\fmf{photon}{i,v1}
\fmf{photon}{v2,o}
\end{fmfgraph*}
\end{gathered}\\
\nonumber
&=&\!2e^2\phi_0\phi^*_0\delta(x-x')g^{\mu\nu}\!+i\Pi^{\mu\nu}_{2,bk}(x,x').
\end{eqnarray}
\end{fmffile}The background plasma responds by particle-hole pair excitation. During this process, a gauge boson is forward scattered, namely, the gauge boson is first absorbed after exciting a plasma particle and then get re-emitted by the particle after its de-excitation. The response due to the vacuum is constituted of the gauge boson mass renormalization and the vacuum polarization 
\begin{fmffile}{vac}
\begin{eqnarray}\label{vac}
\Sigma_{2,vac}^{\mu\nu}(x,x')
&=&\quad
\begin{gathered}
\begin{fmfgraph*}(40,20)
\fmfkeep{hairpin}
\fmfleft{i}
\fmfright{o}
\fmf{photon}{i,v}
\fmf{photon}{v,o}
\fmf{plain}{v,v}
\fmfdot{v}
\fmfv{label=$x$,label.angle=-90,label.dist=6}{v}
\fmfv{label=$\mu$,label.dist=0.2}{i}
\fmfv{label=$\nu$,label.dist=0.2}{o}
\end{fmfgraph*}
\end{gathered}
\quad+\quad
\begin{gathered}
\begin{fmfgraph*}(50,20)
\fmfkeep{bubble}
\fmfleft{i}
\fmfright{o}
\fmf{plain,left=1,tension=0.3}{v1,v2}
\fmf{plain,right=1,tension=0.3}{v1,v2}
\fmfdot{v1,v2}
\fmfv{label=$x$,label.angle=120,label.dist=8}{v1}
\fmfv{label=$x'$,label.angle=60,label.dist=8}{v2}
\fmfv{label=$\mu$,label.dist=0.5}{i}
\fmfv{label=$\nu$,label.dist=0.5}{o}
\fmf{photon}{i,v1}
\fmf{photon}{v2,o}
\end{fmfgraph*}
\end{gathered}\\
\nonumber
&=&2e^2\langle\varphi\varphi*\rangle_{\varphi}\delta(x-x')g^{\mu\nu}+i\Pi^{\mu\nu}_{2,vac}(x,x').
\end{eqnarray}
\end{fmffile}The vacuum responds by virtual pair excitation. During this process, a gauge boson first decays into a pair of virtual particle and antiparticle, and then get reproduced when the virtual pair annihilates. The first line of the effective action (\ref{1-loop}) is the same as $\frac{1}{4}\mathcal{F}_{\mu\nu}\mathcal{F}^{\mu\nu}$ after integration by part. This is the action of the $\mathcal{A}$ field in the vacuum. The second line is a nonlocal term that depends on two coordinates $x$ and $x'$. This term describes the dressing of the $\mathcal{A}$ field due to its interactions with the background plasma and the vacuum.

Explicit expressions of the polarization tensors $\Pi^{\mu\nu}_{2,\text{bk}}(x,x')$ and $\Pi^{\mu\nu}_{2,\text{vac}}(x,x')$ can be found by evaluating the path integrals in (\ref{polarization}). For conciseness, we will abbreviate 1-point functions by $\varphi(x)=\varphi$, $\varphi(x')=\varphi'$, and so on. Similarly, we will abbreviate 2-point functions by $G(x,x')=G$, $G(x,x')=G'$, and so on. It is useful to note the following properties of the path integrals. First, since $\mathcal{L}_\varphi$ is quadratic in $\varphi\varphi^*$, using properties of Gaussian integrals, we know $\langle\varphi\varphi'\rangle_{\varphi}=\langle\varphi^*\varphi'^*\rangle_{\varphi}=0$. Second, due to the imaginary exponential $e^{iS}$, we have $G^*=-G'$. Evaluate the path integrals in (\ref{polarization}) using these properties, we find the background polarization tensor 
\begin{equation}\label{bkPol}
\Pi^{\mu\nu}_{2,\text{bk}}=e^2\big[\phi_0^*\bar{D}^{\mu}-(\bar{D}^{\mu}\phi_{0})^*\big]\big[\phi_0'\bar{D}^{'*\nu}-(\bar{D}^{'\nu}\phi_{0}')\big]G-c.c. \hspace{2pt},
\end{equation}
and the vacuum polarization tensor 
\begin{equation}\label{vacPol}
\Pi^{\mu\nu}_{2,\text{vac}}=e^2\big[G'\bar{D}^{\mu}-(\bar{D}^{*\mu}G')\big](\bar{D}^{'*\nu}G)+c.c.\hspace{2pt}.
\end{equation}
Expressions (\ref{1-loop})-(\ref{vacPol}) combined give an explicit formula of the effective action of gauge field propagation to order $e^2$ in the most general setting. Since the $e^2$-order effective action contains Feynman diagrams up to the 1-loop level, the $e^2$-order effective action is the same as the 1-loop effective action.

Having obtained the formula of the 1-loop effective action, it is worth pointing out a number of symmetries and conservation properties that the effective action has. First, the 1-loop effective action is manifestly Lorentz invariant. Second, it is easy to check that the 1-loop effective action is invariant under the background U(1)-gauge transformation (\ref{BackgroundGauge}), under which the Green's function $G(x,x')$ is transformed by
\begin{equation}\label{eq:GaugeGreen}
G(x,x')\rightarrow e^{ie\chi(x)}G(x,x')e^{-ie\chi(x')}.
\end{equation} 
This property guarantees that the effective action is independent of the choice of the background gauge. So one is free to choose a background gauge that is convenient for calculations. Finally, the 1-loop effective action is invariant under local gauge transformation of the fluctuating gauge field $\mathcal{A}_{\mu}\rightarrow\mathcal{A}_{\mu}+\partial_{\mu}\chi$. Use the Schwinger-Dyson equation of the Green's function
\begin{equation}\label{eq:SDGreen}
[\bar{D}_{\mu}(x)\bar{D}^{\mu}(x)+m^2]G(x,x')=-i\delta(x-x'),
\end{equation}
together with properties of the $\delta-$function
\begin{eqnarray}
f(x)\partial_{\mu}\delta(x)&=&-\delta(x)\partial_{\mu}f(x),\\
\nonumber
f(x,x')\partial_{\mu}\delta(x-x')&=&\frac{1}{2}\delta(x-x')(\partial^{'}_{\mu}-\partial_{\mu})f(x,x'),
\end{eqnarray} 
and the background consistency equations (\ref{EOM}), one can show by straightforward calculations that
\begin{eqnarray}
\label{bkConservation}
\partial_{\mu}\Sigma_{2,\text{bk}}^{\mu\nu}(x,x')&=&\partial'_{\nu}\Sigma_{2,\text{bk}}^{\mu\nu}(x,x')=0,\\
\label{vacConservation}
\partial_{\mu}\Sigma_{2,\text{vac}}^{\mu\nu}(x,x')&=&\partial'_{\nu}\Sigma_{2,\text{vac}}^{\mu\nu}(x,x')=0.
\end{eqnarray}
After integration by part, it is clear that the effective action is invariant under the local gauge transformation of the $\mathcal{A}$ field. Identities (\ref{bkConservation}) and (\ref{vacConservation}) indicate that the plasma current and the vacuum current are conserved separately, so the plasma contribution to wave propagation is separable from the vacuum contribution.

In the above discussion, we obtain the effective action in the configuration space. Sometimes it is more convenient to work in the momentum space, which is related to the configuration space by Fourier transforms
\begin{eqnarray}
\mathcal{A}_{\mu}(x)&=&\int\frac{d^4k}{(2\pi)^4}e^{-ikx}\hat{\mathcal{A}}_{\mu}(k),\\
\label{Fourier}
\Sigma_{2}^{\mu\nu}(x,x')&=&\int\frac{d^4k}{(2\pi)^4}\frac{d^4k'}{(2\pi)^4}e^{-ikx}\hat{\Sigma}_{2}^{\mu\nu}(k,k')e^{ik'x'}_{\hspace{15pt}.}
\end{eqnarray}
The configuration space reality condition $\mathcal{A}^{*}(x)=\mathcal{A}(x)$ and the exchange symmetry $\Sigma_{2}^{\mu\nu}(x,x')=\Sigma_{2}^{\nu\mu}(x,x')$ correspond to the momentum space conditions
\begin{eqnarray}
\hat{\mathcal{A}}_{\mu}(k)&=&\hat{\mathcal{A}}^{*}_{\mu}(-k),\\
\label{eq:reality}
\hat{\Sigma}_{2}^{\mu\nu}(k,k')&=&\hat{\Sigma}_{2}^{\nu\mu}(-k',-k).
\end{eqnarray}
In Fourier space, the $e^2$-order effective action
\begin{eqnarray}\label{MS1Loop}
\nonumber
\Gamma_{2}\bm{[}\mathcal{A}\bm{]}&=&\frac{1}{2}\int\frac{d^4k}{(2\pi)^4} \Bigg(\hat{\mathcal{A}}_{\mu}(-k)(k^{\mu}k^{\nu}-k^2g^{\mu\nu})\hat{\mathcal{A}}_{\nu}(k)\\
&&+\int\frac{d^4k'}{(2\pi)^4}\hat{\mathcal{A}}_{\mu}(-k)\Sigma_2^{\mu\nu}(k,k') \hat{\mathcal{A}}_{\nu}(k')\Bigg),
\end{eqnarray}
where $k^2=k^{\mu}k_{\mu}$ is the Minkowski inner product.

Simplifications can be made when the plasma is translational invariant. In this case, $\Sigma(x,x')$ only depends on the difference between coordinates $r=x-x'$ and is independent of $R=(x+x')/2$. Change variables from $x$ and $x'$ to $r$ and $R$ in Eq. (\ref{Fourier}), we have $\hat{\Sigma}(k,k')=(2\pi)^4\delta^{(4)}(k-k')\hat{\Sigma}(k)$, where $\hat{\Sigma}^{\mu\nu}(k)=\int d^4re^{ikr}\Sigma^{\mu\nu}(r)=\hat{\Sigma}^{\nu\mu}(-k)$. The gauge invariance (\ref{bkConservation}) and current conservation (\ref{vacConservation}) reduce to the Ward$–-$Takahashi identities
\begin{equation}\label{eq:Ward}
k_{\mu}\Sigma_{2,\text{bk}}^{\mu\nu}(k)=k_{\mu}\Sigma_{2,\text{vac}}^{\mu\nu}(k)=0.
\end{equation} 
With the extra delta function, the momentum space $e^2$-order effective action (\ref{MS1Loop}) can be simplified as
\begin{equation}
\Gamma_{2}\bm{[}\mathcal{A}\bm{]}=\frac{1}{2}\!\int\!\frac{d^4k}{(2\pi)^4} \hat{\mathcal{A}}_{\mu}(-k)\Lambda^{\mu\nu}(k)\hat{\mathcal{A}}_{\nu}(k),
\end{equation}
where the dispersion tensor
\begin{equation}\label{eq:DispersionTensor}
\Lambda^{\mu\nu}(k)=k^{\mu}k^{\nu}-k^2g^{\mu\nu}+\Sigma_2^{\mu\nu}(k)=\Lambda^{\nu\mu}(-k).
\end{equation}
For given background fields $\phi_0$ and $\bar{A}$, the dispersion tensor $\Lambda^{\mu\nu}(k)$ is a function of the wave 4-momentum $k_{\mu}$.

\subsection{Observables of effective action}
When there is no external source, the classical equation of motion of $\hat{\mathcal{A}}(k)$ is $\Lambda^{\mu\nu}(k)\hat{\mathcal{A}}_{\nu}(k)=0$. The nontrivial solutions are plane waves whose wave 4-momentum $k$ satisfies $\det\Lambda^{\mu\nu}(k)=0$. The property $k_{\mu}\Lambda^{\mu\nu}(k)=0$ guarantees that one eigenvalue of $\Lambda^{\mu\nu}$ is trivial. 
In fact, using the Ward$–-$Takahashi identity and performing elementary row and column operations, it is easy to show that the temporal components of the dispersion tensor (\ref{eq:DispersionTensor}) can be eliminated by matrix similarity. Hence, the wave dispersion relations are given by 
\begin{equation}\label{dispersion}
\det\Lambda_{ij}(k)=0,
\end{equation}
where $\Lambda_{ij}$ is the spatial block of the dispersion tensor. In general, the 3-by-3 matrix $\Lambda_{ij}$ has three nontrivial eigenvalues, giving relativistic covariant dispersion relations of three waves.

When there is some external test current $\hat{\mathcal{J}}^{\mu}_{\text{ext}}(k)$, the equation of $\hat{\mathcal{A}}(k)$ is $\Lambda^{\mu\nu}(k)\hat{\mathcal{A}}_{\nu}(k)=\hat{\mathcal{J}}_{\text{ext}}^{\mu}(k)$. After gauge fixing, the solution to this inhomogeneous equation is of the form
\begin{equation}
\hat{\mathcal{A}}=\Lambda^{-1}\hat{\mathcal{J}}_{\text{ext}}.
\end{equation}
Taking inverse Fourier transform, the linear response of $\mathcal{A}(x)$ to the external test current $\mathcal{J}_{\text{ext}}(x)$ can be found. For example, when placing a test charge in the plasma $\mathcal{J}^{\mu}_{\text{ext}}(x)=e\delta^{(3)}(\mathbf{x})(1,0,0,0)$, one can derive screening in the relativistic quantum plasma. 

Finally, it is worth pointing out that the configuration space response tensor $\Sigma=\Sigma_{r}+i\Sigma_{i}$ is in general complex, corresponding to the momentum space response tensor $\hat{\Sigma}=\hat{\Sigma}_{H}+i\hat{\Sigma}_{A}$ that contains an antihermitian part. In classical field theory, when one solves the dispersion relation (\ref{dispersion}) with $\Sigma_i\ne 0$, the wave 4-momentum $k^{\mu}$ is necessarily complex. So the amplitude of a plane wave either changes in time in an initial value problem, or changes in space in a boundary value problem. In the quantized field theory, the wave 4-momentum $k^{\mu}$ is always real, and it is the number of gauge bosons that change when $\Sigma_i\ne 0$. By the famous optical theorem, the imaginary part $\Sigma_{i}$ is proportional to the total cross section of the gauge boson. In fact, the optical theorem can be derived from the action exponential as follows. In the configuration space, we can separate the exponentiated action into an oscillatory part and an exponential part:
\begin{equation}
e^{i\Gamma}=e^{iA(\nabla+\Sigma)A}=e^{iA(\nabla+\Sigma_r)A}e^{-A\Sigma_iA}.
\end{equation}
When $\Sigma_i=0$, the exponential is purely oscillatory. This corresponds to the simple propagation of the gauge field. When $\Sigma_i>0$, namely, when the matrix is positive definite, the exponential decays. This corresponds to wave damping in the classical theory, and the decay or absorption of gauge bosons in the quantized theory. When $\Sigma_i<0$, namely, when the matrix is negative definite, the exponential grows. This corresponds to instabilities in the classical theory, and the production or emission of gauge bosons in the quantized theory. Finally, when the matrix $\Sigma_i$ is indefinite, some eigenmodes grow while others decay. In this case, a state of $\mathcal{A}$ field can convert from one mode to another mode as it propagates.

\section{\label{sec:unmagnetized}Wave Propagation in Uniform Unmagnetized Plasmas}
\subsection{Background functions}
When there is no background EM field, it is convenient to choose the vacuum gauge
\begin{equation}
\bar{A}=0
\end{equation} 
In this case, the equation of motion of $\phi_0$ reduces to the Klein-Gordon equation in its simplest form. It is well-known that the single-boson solution to the Klein-Gordon equation is plane waves with the dispersion relation $p^2=p_{\mu}p^{\mu}=m^2$. Since particles are not confined, the background wave functions are not square integrable. To deal with an infinitely large plasma with finite density, it is helpful to first think of a periodic spatial box with size $L$ and a temporal box of length $T$ that contains $N$ particles, and then take the limit that $L, T\rightarrow\infty$ while keeping the density $n_0=N/L^3$ fixed. Inside the box, the properly normalized single-boson wave function 
\begin{equation}
\psi_{\mathbf{p}}^{\epsilon}(x)=\frac{e^{i\epsilon px}}{\sqrt{2mL^3}},
\end{equation}
where $p^{\mu}=(p^0,\mathbf{p})$ is the 4-momentum with $p^0=\sqrt{\mathbf{p}^2+m^2}$. The wave function represents a particle state when $\epsilon=+1$ and an anti-particle state when $\epsilon=-1$. The wave function is normalized such that the current density $\bar{J}_{0}^{\mu}=\epsilon ep^{\mu}/mL^3$ is what one would expect of a single particle. In the periodic box, $\mathbf{p}_{\mathbf{n}}=2\pi\mathbf{n}/L$ is quantized. We can label single particle states by their wave numbers $\bm{n}$. It is clear that the inner products $\langle\psi^{+}_{\bm{n}}|\psi^{+}_{\bm{n}'}\rangle =\langle\psi^{-}_{\bm{n}}|\psi^{-}_{\bm{n}'}\rangle =\delta_{\mathbf{n},\mathbf{n'}}T/2m$ and $\langle\psi^{+}_{\bm{n}}|\psi^{-}_{\bm{n}'}\rangle=0$. 

To see how to deal with an infinitely large plasma, let us calculate current density of $N$ bosons contained in a box. Since particles in plasmas are unbound, they interact weakly with each other. To lowest order, the plasma may be approximated as a collection of noninteracting particles. Suppose the $N$ bosons occupy $M$ orthogonal states $\psi_1,\dots,\psi_M$, with $N_k$ bosons in the state $k$, then the properly symmetrized and normalized wave function
\begin{equation}
\Phi_0=\sqrt{\frac{(2m/T)^{N-1}}{(N-1)!N_1!..N_M!}}\sum_{\sigma\in S_N}\prod_{k=1}^{N}\psi_{d_{\sigma(k)}}(x_k).
\end{equation} 
Here $S_N$ is the permutation group of N-elements. The index function $d_k$ is defined such that $d_k=1$ for $k=1,\dots,N_1$; $d_k=2$ for $k=N_1+1,\dots,N_1+N_2$; and $d_k=M$ for $k=N-N_M+1,\dots,N$. After carrying out the integrals and summations, the current density (\ref{MBcurrent}) becomes $\bar{J}_{0}^{\mu}=e\sum_{k=1}^{M}N_k\epsilon_kp^{\mu}_{k}/mL^3$. More elaborately, the current density can be written as
\begin{equation}
\bar{J}_{0}^{\mu}(x)=\sum_{\epsilon,k}\frac{\rho^{\epsilon}_{k}}{2m}\frac{e}{i}\Big(e^{-i\epsilon_kp_kx}\partial^{\mu}e^{i\epsilon_kp_kx}-c.c\Big),
\end{equation}
where $\rho^{\epsilon}_{k}=N_k/L^3$ is the density of the state $(\epsilon,k)$, and the summation runs over all single-boson states. When $L\rightarrow\infty$, the spectrum of $\phi_0$ becomes continuous. In this case, let a single-boson state be labeled by its wave vector $\mathbf{p}$. If we keep the density of state $\rho^{\epsilon}(\mathbf{p})$ fixed when we take the limit $L,T\rightarrow\infty$, the current density can be written as
\begin{equation}\label{SBcurrent}
\bar{J}_{0}^{\mu}(x)=\sum_{\epsilon=\pm 1}\int\frac{d^3\mathbf{p}}{(2\pi)^3}\frac{e}{i} \Big(\Psi^{\epsilon*}_{\mathbf{p}}(x)\partial^{\mu}\Psi^{\epsilon}_{\mathbf{p}}(x)-c.c\Big),
\end{equation}
where the properly normalized effective single-boson wave function
\begin{equation}\label{Single}
\Psi_{\mathbf{p}}^{\epsilon}(x)=\sqrt{\frac{\rho^{\epsilon}(\mathbf{p})}{2m}}e^{i\epsilon px}.
\end{equation}
We see that the current (\ref{SBcurrent}) can be obtained from the many-body current (\ref{MBcurrent}) by replacing $\phi_0(x)$ with the properly normalized effective single-boson wave function (\ref{Single}), followed by summations and integrations over the Hilbert space of single-boson states.

Besides the wave functions, we will also need the Green's function of the free $\varphi$ field. It is well-known that when $\bar{A}=0$, the Green's function
\begin{equation}\label{eq:UMGreen}
G(x,x')=\int\frac{d^4k}{(2\pi)^4}\frac{ie^{-ik(x-x')}}{k^2-m^2}.
\end{equation} 
For path integrals to converge, one needs to replace $m^2\rightarrow m^2-i0$ in the exponentiated action. This gives a prescription of the integration contour around poles in the complex plane when evaluating the integral in Eq. (\ref{eq:UMGreen}). The resultant Green's function is the Feynman Green's function. 

\subsection{Vacuum response}
The vacuum response when $\bar{A}=0$ is also well-known. In Fourier space, the vacuum response tensor
\begin{equation}
\hat{\Sigma}_{2,\text{vac}}^{\mu\nu}(k)=\chi_{v}(k^2)(k^{\mu}k^{\nu}-k^2g^{\mu\nu}).
\end{equation}
The Ward$–-$Takahashi identity is manifestly satisfied. Using dimensional regularization in $d<4$ dimensions for divergent 1-loop integrals, imposing the renormalization condition that photons remain massless in the vacuum, and subtracting the counter terms, we obtain the renormalized vacuum permittivity
\begin{eqnarray}\label{UMvac}
\nonumber
\chi_{v}(k^2)&=&e^2\frac{\Gamma(2-d/2)}{(4\pi)^{d/2}}\int_{0}^{1}du\frac{(1-2u)^2}{(m^2)^{2-d/2}}\\
\nonumber
&&\times\Big[\Big(1-u(1-u)k^2/m^2\Big)^{d/2-2}-1\Big]\\
\nonumber
&=&\frac{2e^2}{3(4\pi)^2}\Big\{\frac{4}{3}-\frac{4m^2}{k^2}+\Big(\frac{4m^2}{k^2}-1\Big)^{3/2}\\
&&\times\arctan\Big[\Big(\frac{4m^2}{k^2}-1\Big)^{-1/2}\Big]\Big\}.
\end{eqnarray}
On the first line, $\Gamma(z)$ is the gamma function, which diverges when $z$ equals to nonpositive integers. To obtain the second equality, the limit $d\rightarrow 4$ is taken, followed by integration over the Feynman parameter $u$. It is not hard to see that $\hat{\Sigma}_{2,\text{vac}}(k^2)$ is real when $k^2=k_{\mu}k^{\mu}\le 4m^2$, and $\hat{\Sigma}_{2,\text{vac}}(k^2)$ becomes complex with a positive imaginary part when $k^2>4m^2$. The positive imaginary part is proportional to the cross section of a gauge boson, which can decay into a pair of charged particle and antiparticle when $k^2>4m^2$.

\subsection{Plasma Response}
The plasma response $\Sigma_{2,\text{bk}}^{\mu\nu}(x,x')$ can be evaluated by substituting the effective single-boson wave function (\ref{Single}) and the Green's function (\ref{eq:UMGreen}) into Eq. (\ref{bkPol}), followed by integration and summation over the single-boson Hilbert space. The contribution of each charged species to the mass term of the $\mathcal{A}$ field is
\begin{equation}\label{eq:bkmass}
2e^2\phi_0\phi^*_0=\sum_{\epsilon=\pm 1}\int\frac{d^3\mathbf{p}}{(2\pi)^3}\frac{e^2\rho^{\epsilon}(\mathbf{p})}{m},
\end{equation}
and the plasma polarization tensor (\ref{bkPol}) becomes
\begin{eqnarray}\label{eq:bkPol}
\nonumber
\Pi^{\mu\nu}_{2,\text{bk}}&=&\sum_{\epsilon=\pm 1}\int\frac{d^3\mathbf{p}}{(2\pi)^3} e^2[\Psi_{\mathbf{p}}^{\epsilon*}\partial^{\mu}-(\partial^{\mu}\Psi_{\mathbf{p}}^{\epsilon})^*]\\
&&\qquad\times[\Psi_{\mathbf{p}}^{\epsilon'}\partial^{'\nu}-(\partial^{'\nu}\Psi_{\mathbf{p}}^{\epsilon'})]G-c.c.\hspace{2pt}.
\end{eqnarray}
Similar results are shown by Melrose \cite{Melrose07} using the prescription of cutting one charged particle propagator in the vacuum polarization diagrams and replacing it by statistical average over the plasma. Our formulation has thus provided a justification for such a prescription. 

Let us consider the example of a cold particle plasma. Denote the 4-momentum of plasma particles by $q^{\mu}$, then the density of state 
\begin{equation}\label{UmagDOS}
\rho^{\epsilon}(\mathbf{p})=n_0(2\pi)^3\delta^{(3)}(\mathbf{p}-\mathbf{q})\delta_{\epsilon,1}\hspace{2pt},
\end{equation} 
where $n_0$ is the number density of the plasma. Because of the $\delta$-functions, integrals and summations can be carried out very easily. The current density due to each charged species becomes
\begin{equation}
\bar{J}_{0}^{\mu}(x)=en_0q^{\mu}/m.
\end{equation} 
This is what one would expect of a cold uniform fluid. To satisfy the background self-consistency $\partial_{\mu}\bar{F}^{\mu\nu}=0$, the plasma needs to be constituted of more than one charged species, such that the total current $\bar{J}_{0}^{\mu}=0$ when summed over all charged species. Using the density of state (\ref{UmagDOS}), the mass term of the $\mathcal{A}$ field (\ref{eq:bkmass}) becomes
\begin{equation}\label{UMmass}
2e^2\phi_0(x)\phi^*_0(x)=\frac{e^2n_{0}}{m}=\omega_{p}^2.
\end{equation}
It is easy to recognize that $\omega_p$ is the plasma frequency. The plasma polarization tensor (\ref{eq:bkPol}) becomes
\begin{eqnarray}\label{UMpol}
&&\Pi^{\mu\nu}_{2,\text{bk}}(x,x')=\frac{\omega_{p}^2}{2}\int\frac{d^4k}{(2\pi)^4}ie^{-ik(x-x')}\\
\nonumber
&&\hspace{35pt}\times\Big(\frac{(2q+k)^{\mu}(2q+k)^{\nu}}{(k+q)^2-m^2} +\frac{(2q-k)^{\mu}(2q-k)^{\nu}}{(k-q)^2-m^2}\Big).
\end{eqnarray} 
The two terms above correspond to the $s$-channel and the $t$-channel Feynman diagrams of the forward scattering of a gauge boson. We see quantum recoil, the change of the 4-momentum of charged particles during forward scattering of gauge bosons, is automatically taken into account. Combining (\ref{UMmass}) with (\ref{UMpol}) and taking Fourier transform, the contribution of each charged species to the momentum space plasma response tensor is
\begin{equation}\label{UMbk}
\hat{\Sigma}_{2,\text{bk}}^{\mu\nu}=\omega_{p}^2\bigg(g^{\mu\nu}-\frac{k^2(4q^{\mu}q^{\nu}\!+k^{\mu}k^{\nu}) -4kq(q^{\mu}k^{\nu}\!+k^{\mu}q^{\nu})}{(k^2)^2-4(kq)^2}\bigg).
\end{equation}
Here $k^2=k^{\mu}k_{\mu}$ and $kq=k^{\mu}q_{\mu}$ are Minkowski inner products. It is straightforward to check that the Ward$–-$Takahashi identity is satisfied.

\subsection{Wave dispersion relations}
Having found the response tensors due to the vacuum (\ref{UMvac}) and the cold plasma (\ref{UMbk}), we can proceed to find dispersion relations. The simplest case is when different charged species in the plasma have no relative motion. In this case, there is an inertial frame in which all background particles are at rest. In this plasma rest frame, the particle 4-momentum $q^{\mu}=(m,0,0,0)$. Let us choose a coordinate system such that wave 4-momentum $k^{\mu}=(\omega,\mathrm{k},0,0)$. In this coordinate system, the nonzero components of the plasma response tensor are\footnote{We use the italics $k$ for 4-momentum of and the roman $\mathrm{k}=|\bm{k}|$ for the magnitude of the wave vector. While $k^2=k^{\mu}k_{\mu}$ and $kq=k^{\mu}q_{\mu}$ denote the Minkowski inner products, $\mathrm{k}^2$ and $\omega\mathrm{k}$ are products of scalars.}
\begin{eqnarray}
\nonumber
\hat{\Sigma}_{2,\text{bk}}^{00}&=&\chi_p \mathrm{k}^2,\\
\nonumber
\hat{\Sigma}_{2,\text{bk}}^{11}&=&\chi_p \omega^2,\\
\nonumber
\hat{\Sigma}_{2,\text{bk}}^{01}&=&\hat{\Sigma}_{2,\text{bk}}^{10}=\chi_p\omega \mathrm{k},\\
\hat{\Sigma}_{2,\text{bk}}^{22}&=&\hat{\Sigma}_{2,\text{bk}}^{33}=-\omega_{p}^{2},
\end{eqnarray} 
where
\begin{eqnarray}
\nonumber
\omega_{p}^{2}&=&\sum_{s}\omega_{ps}^{2},\\
\chi_p&=&\sum_{s}\frac{\omega_{ps}^{2}(\bm{k}^2-\omega^2+4m_{s}^2)}{(\omega^2-\bm{k}^2)^2-4m_{s}^2\omega^2},
\end{eqnarray}
are the plasma frequency and the plasma susceptibility, respectively. 

Using elementary column and row operations, the dispersion matrix $\Lambda^{\mu\nu}$ can be diagonalized and the eigenvalue problem can be solved. There are two transverse modes and one longitudinal mode. The two transverse modes are degenerate and electromagnetic with the dispersion relation
\begin{equation}\label{EM}
(1+\chi_v)(\omega^2-\bm{k}^2)-\omega_{p}^{2}=0.
\end{equation}
From this dispersion relation, it is easy to see that the photon modes are gapped when background plasmas exist. Namely, the wave frequency $\omega\neq 0$ when the wave vector $\mathrm{k}=0$ if $\omega_p\neq 0$ . The longitudinal mode is electrostatic with the dispersion relation
\begin{equation}\label{langmuir}
1+\chi_v+\chi_p=0.
\end{equation}
Since $\chi_p(\bm{k}=0)=-\omega_{p}^{2}/\omega^2$, there always exist one gapped plasmon mode, known classically as the Langmuir wave. When there are two or more charged species, there also exists a gapless phonon mode, known classically as the ion acoustic wave. 

\begin{figure}[h]
	\renewcommand{\figurename}{FIG.}
	\includegraphics[angle=0,width=8.6cm]{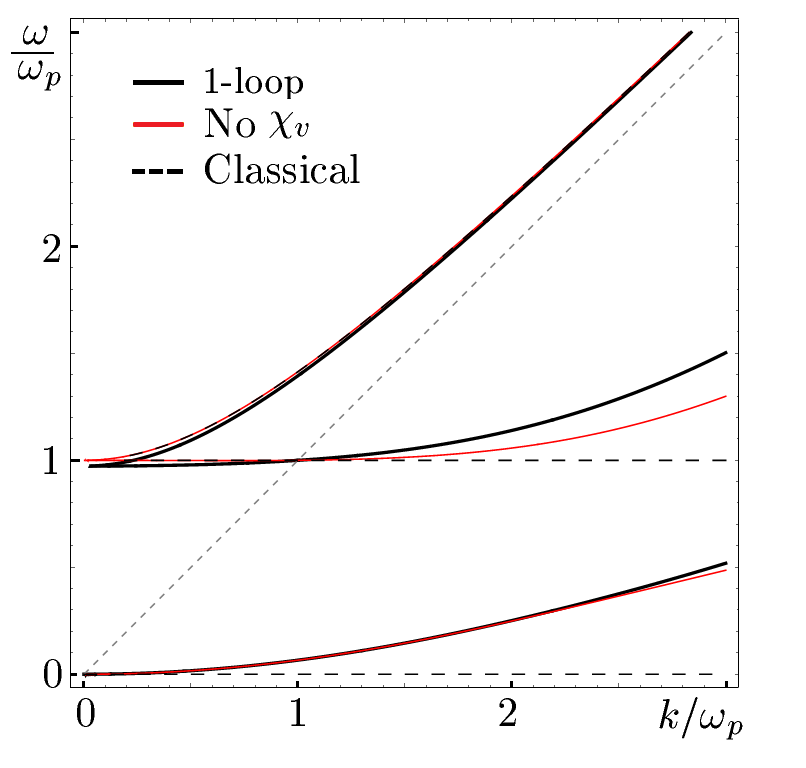}
	\caption{Wave dispersion relations in a cold, unmagnetized, quasineutral, spinless, electron-ion plasma. For various effects to be visible on the scale of this figure, parameters used for making this plot are $2e^2/3(4\pi)^2=20$, $m_e/\omega_{pe}=5$, and $m_i/m_e=3$. The solid black curves are the 1-loop dispersion relations. The solid red curves are the dispersion relations that ignore the vacuum polarization. The dashed black curves are wave dispersion relations in a classical plasma. The upper curves are the electromagnetic waves, the middle curves are the Langmuir waves, the bottom curves are the ion acoustic waves, and the dashed gray line across the diagonal represents the light cone. Notice that near the light cone, wave dispersion relations in the relativistic quantum plasma asymptote to wave dispersion relations in the classical plasma.}
	\label{fig:UMDispersion}
\end{figure}

An example of wave dispersion relations in a cold, quasineutral, spinless, electron-ion plasma is plotted in Fig. \ref{fig:UMDispersion}. In the figure, the upper curves are the degenerate EM waves, the middle curves are the Langmuir waves, and the lower curves are the ion acoustic waves. There also exist high energy solutions with $\omega>2m$. These high energy modes readily decay and are not plotted here. In Fig. \ref{fig:UMDispersion}, the solid black curves are the 1-loop dispersion relations solved from (\ref{EM}) and (\ref{langmuir}). The red curves are wave dispersion relations when the vacuum polarization $\chi_v$ is ignored. The dashed black curves are wave dispersion relations in a classical plasma. In this example, we set the plasma density high with $\omega_{pe}/m_e=0.2$, such that relativistic quantum effects are visible on the scale of this figure. We set the ions mass low with $m_i/m_e=3$, such that ion effects are comparable to electron effects. The coupling constant is taken to be unphysically strong $2e^2/3(4\pi)^2=20$, such that vacuum effects are comparable to plasma effects on the scale of this figure. As can be seen from the figure, wave dispersion relations in the relativistic quantum plasma are very similar to those in the classical plasma near the light cone.  However, there are clearly distinctions away from the light cone. In particular, unlike classical plasma theories,  the relativistic quantum theory predicts that longitudinal waves propagate with nonzero group velocities even when the plasma is cold. This can be understood intuitively. Since the longitudinal wave quanta spend part of their time being carried by some excited charged bosons, they propagate together with these charged bosons, which have nonzero velocities.

The photon modes and the plasmon mode have the same cutoff frequency $\omega_c$, which is given by the solution to the equation $\omega_c^2(1+\chi_v(\omega_c,\mathrm{k}=0))=\omega_{p}^{2}$. The cutoff frequency $\omega_c$, or the mass gap, is less than the plasma frequency $\omega_p$ due to the vacuum polarization. It can be shown that the ratio $\omega_c/\omega_p$ decreases with increasing $\omega_p/2m$. This can be understood intuitively. Vacuum polarization produces virtual pairs near charged particles. These virtual pairs screen the electric field of charged particles, so the effective electric charge of particles are reduced. For higher plasma densities, the virtual pair density is also higher, resulting in stronger shielding of the electric charge and consequently smaller cutoff frequencies. To get a sense of how small the vacuum polarization effect is, let us approximate $\omega_c$ when $\omega_p/2m=1$. Write $g=2e^2/3(4\pi)^2$. Since the physical value for electron charge is $g=\alpha/6\pi\ll 1$, where $\alpha\approx 1/137$ is the fine structure constant, the equation for $\omega_c$ can be solved asymptotically. To lowest order, $\omega_c/\omega_p\sim 1-g/6$. We see the effect of the vacuum polarization is minuscule. 

Now let us check that classical dispersion relations are recovered when taking the classical limit in relativistic quantum results. Since energy of particles are not quantized in unmagnetized plasmas, the non-relativistic low energy limit is the classical limit. In the low energy limit $k^2/m^2\rightarrow 0$, namely, near the light cone, contribution of a relativistic quantum plasma asymptote to that of a classical plasma
\begin{equation}
\chi_p\sim-\sum_{s}\frac{\omega_{ps}^{2}}{\omega^2}\Big(1+\frac{k^2}{4m_{s}^2}\Big)\Big(1+\frac{k^4}{4m_{s}^2\omega^2}\Big) \rightarrow-\frac{\omega_{p}^{2}}{\omega^2},
\end{equation}
and the contribution of the vacuum vanishes
\begin{equation}
\chi_v\sim\sum_{s}\frac{g_s}{5}\frac{k^2}{4m_{s}^2-k^2}\rightarrow 0.
\end{equation}
This can be understood intuitively, since gauge bosons in this limit do not have sufficient energy to create pairs of massive charged particle and anti-particle. In the low energy limit, the asymptotic dispersion relations of the photon, the plasmon, and the phonon modes are
\begin{eqnarray}\label{eq:UMDispApprix}
\nonumber
\omega^2&\sim&\omega_p^2(1-\lambda_D^2\omega_p^2)+\bm{k}^2,\\
\nonumber
\omega^2&\sim&\omega_p^2\Big(1\!-\!\lambda_D^2(\omega_p^2-\bm{k}^2)\Big)\!-\!\Big(\frac{\omega_{pe}^2}{4m_e^2}+\frac{\omega_{pi}^2}{4m_i^2}\Big) \Big(\bm{k}^2-\frac{\bm{k}^4}{\omega_p^2}\Big),\\
\omega^2&\sim&\Big(\frac{\omega_{pe}^2}{4m_i^2}+\frac{\omega_{pi}^2}{4m_e^2}\Big) \frac{\bm{k}^4}{\omega_p^2}=\frac{\bm{k}^4}{4m_em_i}.
\end{eqnarray}
Here $\lambda_D^2=\sum_{s}g_s/20m_s^2 $ is the vacuum shielding length due to virtual pair production. Similar results are obtained in \cite{Hines78, Kowalenko85}, in which the phonon mode was not considered. 

In the opposite limit $k^2/m^2\rightarrow\infty$, namely, away from the light cone, the plasma contribution diminishes
\begin{equation}\label{eq:pasym}
\chi_p\sim-\frac{\omega_{p}^{2}}{k^2}\rightarrow 0.
\end{equation}
This can be understood from the perspective of spatial scales by fixing $\omega$ and letting $\bm{k}$ go to infinity. In this perspective, since the wave length of a high energy gauge boson is much smaller than the typical inter-particle spacing in the plasma, the gauge boson rarely encounters a plasma particle and propagates as if it is in the vacuum. The asymptotic behavior (\ref{eq:pasym}) can also be understood from the perspective of time scales by fixing $\bm{k}$ and letting $\omega$ go to infinity. In this perspective, since the wave frequency is much larger than the plasma frequency, the plasma does not have time to respond. Unlike the plasma response, which diminishes in the high energy limit $k^2/m^2\rightarrow\infty$, the real part of the vacuum susceptibility blows up 
\begin{equation}
\text{Re}(\chi_v)\sim-\sum_{s}\frac{g_{s}}{2}\ln\Big|\frac{k^2}{4m_{s}^2}\Big|\rightarrow\infty,
\end{equation}
This can be understood intuitively. Since the gauge boson has sufficient energy to create massive charged particles in this limit, the vacuum fluctuation is large. Outside the light cone, the imaginary part of $\chi_v$ is always zero. Inside the light cone, when $k^2> 4m^2$, the imaginary part of the vacuum susceptibility 
\begin{equation}
\text{Im}(\chi_v)\sim\sum_{s}\frac{\pi}{2}\Big(1-\frac{4m_{s}^2}{k^2}\Big)\rightarrow\sum_{s}\frac{\pi}{2}, 
\end{equation} 
This positive imaginary part is proportional to the total decay cross section of a massive gauge boson. The imaginary part is larger when there are more charged species, in which case there are more types of particles that the massive gauge boson can decay into. After its typical life time, a massive gauge boson decays and thereafter stops propagating.

\section{\label{sec:magnetized}Wave Propagation in Uniformly Magnetized Plasmas}
\subsection{Background functions}
Let us choose a coordinate system such that the uniform magnetic field with strength $B_0$ points in the $+z$ direction. Since the choice of gauge does not affect final results, we are free to use the symmetric gauge
\begin{equation}\label{eq:SymmetricGauge}
\bar{A}^{\mu}=(0,-\frac{1}{2}B_0y,\frac{1}{2}B_0x,0).
\end{equation}
This gauge is convenient since it respects the rotation symmetry of the system.

Having chosen the background gauge $\bar{A}$, the field $\phi_0$ can be solved from its equation of motion (\ref{EOM}). The solutions are relativistic Landau levels, which have been obtained by a number of authors \cite{Witte87}. Since particles are confined by the magnetic field in the $xy$-plane, for a single-boson wave function to be normalizable, it is only necessary to impose a periodic box in the $z$-direction. Let $L$ be the length of this periodic box. Then in polar coordinate, properly normalized single-boson wave functions are 
\begin{eqnarray}\label{eq:MagSingle}
\nonumber
\psi_{n,l,p_\parallel}^{\epsilon}(x)&=&\sqrt{\frac{n!}{2m_{n}l!\pi r_0^2L}}\Big(\frac{r}{r_0}\Big)^{l-n} \Lambda_{n}^{(l-n)}(r^2/r_0^2)\\
&\times&\exp\bm{\{} i\epsilon[E_{n,p_\parallel}t-p_\parallel z\mp(l-n)\theta]\bm{\}}.
\end{eqnarray}
Here $n=0,1,2\dots$ is the principle quantum number, $l=0,1,2\dots$ is the angular momentum quantum number, $p_\parallel$ is the parallel momentum, and $\epsilon=+1$ and $-1$ for particle and antiparticle states, respectively. The relativistic Landau level with principle quantum number $n$ and parallel momentum $p_\parallel$ has energy
\begin{eqnarray}
E_{n,p_\parallel}&=&\sqrt{m_n^2+p_\parallel^2}\\
\nonumber
&\sim&m+\frac{p_\parallel^2}{2m}+|\Omega|(n+\frac{1}{2}), \quad p_\parallel,|\Omega|\ll m.
\end{eqnarray}
where $\Omega=eB_0/m$ is the gyrofrequency and the effective mass of the $n$-th excited state is
\begin{equation}
m_n=\sqrt{m^2+|eB_0|(2n+1)}.
\end{equation}
In radial wave functions, $\Lambda_{n}^{(\alpha)}(x) =L_{n}^{(\alpha)}(x)e^{-x/2}$ is the generalized Laguerre function, where $L_{n}^{(\alpha)}(x)$ is the generalized Laguerre polynomial. The radial wave functions are localized within a length scale set by the B-field 
\begin{equation}
r_{0}=\sqrt{2/|eB_0|}.
\end{equation}
In the azimuthal wave functions, the upper and lower sign of $\mp$ in front of $\theta$ correspond to the case $eB_0>0$ and $eB_0<0$, respectively. These signs account for the fact that positively and negatively charged particles gyrate in opposite directions. To see the wave functions are properly normalized, one can calculate, for example, the total current in the $z$-direction.

To deal with plasmas that are infinitely large, we follow procedures in Sec. \ref{sec:unmagnetized}. Namely, we first consider finite number of particles in a spatial box of size $L$ and temporal box of length $T$. Using the many-body wave function and taking the limit $L,T\rightarrow\infty$ while keeping the plasma density fixed is equivalent to using the effective single-boson wave function 
\begin{eqnarray}\label{eq:MagSingleEff}
\nonumber
\Psi_{n,l,p_\parallel}^{\epsilon}(x)&=&\sqrt{\frac{n!\rho^{\epsilon}_{n,l}(p_\parallel)}{2m_{n}l!}}\Big(\frac{r}{r_0}\Big)^{l-n} \Lambda_{n}^{(l-n)}(r^2/r_0^2)\\
&\times&\exp{\{ i\epsilon[E_{n,p_\parallel}t-p_\parallel z\mp(l-n)\theta]\}},
\end{eqnarray}
followed by integration over the continuous label $p_\parallel/2\pi$ and summations over the discrete labels $n,l$,and $\epsilon$ over the single-boson Hilbert space. In the above expression, $\rho^{\epsilon}_{n,l}(p_\parallel)$ is again the density of state. 

Besides the wave functions, we will also need the Green's function of charged bosons in a uniform magnetic field. The Green's function can either be found by calculating the propagator of the quantized $\varphi$ field, or more directly by solving the Schwinger$–-$Dyson equation (\ref{eq:SDGreen}). There are many representations of the Green's function, for example, the spectral representation \cite{Melrose12} and the proper time representation \cite{Schwinger51}. A useful expression of the Green's function for our purpose is obtained in Appendix \ref{app:Green}. The Green's function can be put into the form
\begin{eqnarray}\label{eq:MagGreen}
&&G(x,x')=\frac{i}{\pi r_{0}^{2}}e^{ieB_0(xy'-yx')/2}\sum_{n=0}^{\infty}\int\!\frac{dq_0dq_\parallel}{(2\pi)^2}\\
\nonumber
&&\hspace{20pt}\times\frac{e^{i[q_{0}(t-t')-q_{\parallel}(z-z')]}}{q_{0}^{2}-q_{\parallel}^{2}-m_n^2}\Lambda_{n}^{(0)}\bm{\Big(}\frac{(x-x')^2+(y-y')^2}{r_0^2}\bm{\Big)}.
\end{eqnarray}
This form of the Green's function respects a number of symmetries of the system. It is manifestly invariant under the rotation around the $z$-axis. It is also invariant under parity $\mathbf{x}\rightarrow-\mathbf{x}$. Finally, it is invariant under the joint symmetry action of charge conjugation $e\rightarrow-e$ and time reversal $t\rightarrow-t, B_0\rightarrow-B_0$. These demanded symmetry properties will enable easy extraction of physics later on. Notice that the Green's function has poles when $q_0=\pm E_{n,q_\parallel}$. These are nothing other than the dispersion relations of charged particles occupying relativistic Landau levels. When the $\varphi$ field propagates, it can propagate through any of these channels.

\subsection{Vacuum response}
We will not consider the vacuum response in this section for three reasons. First, as can be seen from Sec. \ref{sec:unmagnetized}, effects of the vacuum response are minuscule compared to effects of the plasma response for low energy waves\footnote{It is safe to ignore the vacuum response when the wave rest energy $m_\gamma c^2=\sqrt{k^{\mu}k_{\mu}}c\hbar$ and the cyclotron energy $\hbar\Omega$ are smaller than the electron rest energy $m_ec^2$.}. Second, due to (\ref{bkConservation}) and (\ref{vacConservation}), contributions by plasmas and the vacuum are separable. Ignoring the vacuum response does not break any symmetry of the system. Finally, for a practical reason, obtaining a useful expression of the vacuum response is highly nontrivial. Although many representations of the response tensor have been obtained \cite{Witte90}, they can be evaluated analytically only in some special limits \cite{Karbstein13, Shabad75}.

\subsection{Plasma response}
After choosing the density of state appropriately, such that the equation for the background EM field $\partial_{\mu}\bar{F}^{\mu\nu}=0$ is satisfied, we can evaluate the plasma response tensor ({\ref{bkPol}}) by plugging in the effective single-boson wave functions (\ref{eq:MagSingleEff}) and the Green's function (\ref{eq:MagGreen}), followed by integration over the continuous label $p_\parallel/2\pi$ and summations over the discrete labels $n,l$, and $\epsilon$. 

Let us consider the example of a cold particle plasma, in which all charged bosons are condensed in the lowest Landau levels. In the rest frame of the cold plasma, the density of state 
\begin{equation}\label{eq:MagDOS}
\rho^{\epsilon}_{n,l}(p_\parallel)=2\pi n_0\delta(p_\parallel)\delta_{n,0}\delta_{\epsilon,1}\hspace{2pt},
\end{equation}
where $n_0$ is the number density of the plasma. The three $\delta$-functions make it very easy to carry out the integrations over $p_\parallel/2\pi$ and summations over $n,l$, and $\epsilon$. Using covariant derivatives listed in Appendix \ref{app:Derivative}, it is easy to calculate the many-body current. The current density due to each charged species is
\begin{equation}
\bar{J}_{0}^{\mu}=en_0(1,0,0,0).
\end{equation}
This is what one would expect of a uniform cold fluid. The self-consistency condition $\sum_s\bar{J}_{s0}^{\mu}=0$ is satisfied if the plasma is quasineutral. The contribution of each charged species to the mass term of the $\mathcal{A}$ field is 
\begin{equation}
2e^2\phi_0\phi_0^*=\frac{m\omega_p^2}{m_0},
\end{equation}
where $\omega_p^2=e^2n_0/m$ is again the plasma frequency. The calculation of the plasma polarization tensor is more technical. We show details of the calculation of $\Pi_{2,\text{bk}}^{00}$ in Appendix \ref{app:polarization}. Other components of the polarization tensor can be calculated using similar methods. Combining the polarization term with the mass term, the contribution of each charged species to the Fourier space plasma response tensor is
\begin{eqnarray}\label{MagPol}
\nonumber
\hat{\Sigma}^{\lambda\sigma}_{2,\text{bk}}(k)&=&\frac{m\omega_p^2}{m_0}\bigg(g^{\lambda\sigma}-\frac{1}{2}\sum_{\varsigma=\pm 1}(\kappa+\varsigma\varrho)^{\lambda}(\kappa+\varsigma\varrho)^{\sigma}K_\varsigma^{(0)}\bigg),\\
\nonumber
\hat{\Sigma}^{ab}_{2,\text{bk}}(k)&=&\frac{m\omega_p^2}{2m_0}\sum_{\varsigma=\pm 1}\bigg(\varepsilon^{ac}\varepsilon^{bd}\kappa^c\kappa^d(2K_\varsigma^{(1)}-K_\varsigma^{(0)})\\
\nonumber
&&-\kappa_\varsigma^2[\delta^{ab}K_\varsigma^{(1)}\pm i\varsigma\varepsilon^{ab}(K_\varsigma^{(1)}-K_\varsigma^{(0)})]\bigg),\\
\nonumber
\hat{\Sigma}^{\lambda a}_{2,\text{bk}}(k) &=&\hat{\Sigma}^{a\lambda}_{2,\text{bk}}(-k)=\frac{m\omega_p^2}{2m_0}\sum_{\varsigma=\pm 1}(\kappa+\varsigma\varrho)^{\lambda}\\
&&\times\bigg(-\kappa^aK_\varsigma^{(1)}\pm i\varsigma\varepsilon^{ab}\kappa^b(K_\varsigma^{(1)}-K_\varsigma^{(0)})\bigg).
\end{eqnarray}
In the above expressions, the Greek indices $\lambda,\sigma=0,3$ correspond to the unconfined directions, and the Latin indices $a,b=1,2$ correspond to the confined directions. On right hand sides, $g^{\lambda\sigma}$ is the metric tensor of the Minkowsi space, $\delta^{ab}$ is the $\delta$-function, and $\varepsilon^{ab}$ is the 2-dimensional Levi-Civita symbol. The upper and lower sign of $\pm$ in the imaginary parts correspond the the case $eB_0>0$ and $eB_0<0$, respectively. For conciseness, we denote $K_\varsigma^{(n)}:= K(\kappa_\varsigma^2-n,\bm{\kappa}^2)$, where $\kappa_\varsigma^2:=\kappa_0^2-\kappa_3^2+\varsigma\varrho_0\kappa_0$,  $\bm{\kappa}^2:=\kappa_1^2+\kappa_2^2$, and the $K$-function is related to the confluent hypergeometric function ${}_1F_1(a;b;z)$ by 
\begin{eqnarray}\label{eq:Kfunction}
K(x,z)&:=&\frac{1}{x}{}_1F_1(1;1-x;-z).
\end{eqnarray}
The wave 4-momentum is normalized by the magnetic field length $\kappa^{\mu}=r_0k^{\mu}/2$ and the 4-momentum of the plasma particle is normalized by $\varrho^{\mu}=r_0(m_0,0,0,0)$. The summation over $\varsigma=\pm 1$ corresponds to the summation of the $s$-channel and the $t$-channel Feynman diagrams of the forward scattering of a gauge boson.

The plasma response tensor (\ref{MagPol}) satisfies a number of required symmetries. It satisfies the exchange symmetry $\hat{\Sigma}^{\mu\nu}_{2,\text{bk}}(k)=\hat{\Sigma}^{\nu\mu}_{2,\text{bk}}(-k)$, as required by (\ref{eq:reality}). It is invariant under the rotation around the $z$-axis, which is a basic symmetry of the system. Moreover, it transforms properly under time reversal symmetry $T^{\mu}_{\nu}=\text{diag}(-1,1,1,1)$ by $\hat{\Sigma}^{\mu\nu}(\omega,\bm{k})|_{B_0}=T^{\mu}_{\alpha}T^{\nu}_{\beta}\hat{\Sigma}^{\alpha\beta}(-\omega,\bm{k})|_{-B_0}$. Finally, using property (\ref{eq:Kconfluent}) of the confluent hypergeometric function, it is straightforward to check that the Ward$–-$Takahashi identity (\ref{eq:Ward}), which is required by charge conservation and gauge invariance, is satisfied. In addition to these symmetry properties, the plasma response tensor (\ref{MagPol}) has a number of asymptotic properties. First, since the confluent hypergeometric function ${}_1F_1(a;b;z)$ has poles whenever $b$ equals to nonpositive integers, the response tensor has poles whenever $\omega=\pm\omega_{n,k_\parallel}^{\pm}$, where the frequency of relativistic quantum cyclotron resonances
\begin{eqnarray}\label{eq:Poles}
\omega_{n,k_\parallel}^{\pm}&=&E_{n,k_\parallel}\pm m_0\\
\nonumber
&\sim& \left\{ \begin{array}{ll}
\omega_{n,k_\parallel}^{-}+2(m+|\Omega|/2), & ``+",\\
k_\parallel^2/2m+n|\Omega|, & ``-".
\end{array} \right.
\end{eqnarray}
The above asymptotic behavior is in the limit $k_\parallel, |\Omega|\ll m$. These resonances have clear physical meanings. The $\omega_{n,k_\parallel}^{-}$ resonance corresponds to the energy it takes to excite a plasma particle from the ground state to the $n$-th Landau level with parallel momentum $k_\parallel$. The $\omega_{n,k_\parallel}^{+}$ resonance corresponds to the aforementioned excitation energy plus the energy it takes to create a pair of new particles in ground states. The second important asymptotic property of the response tensor (\ref{MagPol}) is when the magnetic field $B_0\rightarrow 0$. In this limit, the ground state mass $m_0$ asymptotes to the bare mass $m$. Moreover, use the asymptotic property that ${}_1F_1(1;b;z)\rightarrow b/(b-z)$ when $z,b\rightarrow\infty$ while keeping $b/z$ fixed \cite{NIST10}, we can find the asymptotic behavior of the $K$-function 
\begin{equation}
\frac{r_0^2}{4}K_{\epsilon}^{(n)}\rightarrow\frac{1}{k^2+2\epsilon m k_0}, \qquad B_0\rightarrow 0.
\end{equation}
Here $k^2=k^{\mu}k_{\mu}$ is the Minkowski inner product. Using the above expression, it is straightforward to check that in the limit $B_0\rightarrow 0$, the response tensor (\ref{MagPol}) of cold magnetized plasmas asymptotes to the response tensor (\ref{UMbk}) of cold unmagnetized plasmas.

\subsection{Wave dispersion relations}
Let us choose a coordinate system such that the wave 4-momentum $k^{\mu}=(\omega,k_\perp,0,k_\parallel)$. Since $k_2=0$ in this coordinate system, the plasma response tensor (\ref{MagPol}) can be simplified. To facilitate discussions next, let us write the wave dispersion relation (\ref{dispersion}) explicitly 
\begin{equation}\label{eq:MagDisp}
\det\!\left(\!\begin{array}{ccc}
\omega^2\!-\!k_{\parallel}^2\!+\!\hat{\Sigma}^{11}\!&\!\hat{\Sigma}^{12}\!&\!k_\perp k_\parallel\!+\!\hat{\Sigma}^{13}\!\\
\hat{\Sigma}^{21}\!&\!\omega^2\!-\!\bm{k}^2\!+\!\hat{\Sigma}^{22}\!&\!\hat{\Sigma}^{23}\!\\
k_\perp k_\parallel\!+\!\hat{\Sigma}^{31}\!&\!\hat{\Sigma}^{32}\!&\!\omega^2\!-\!k_\perp^2\!+\!\hat{\Sigma}^{33}\!
\end{array}\!\right)\!=\!0.
\end{equation}
The situation becomes particularly simple when the wave vector is exactly parallel ($k_\perp=0$) or perpendicular ($k_\parallel=0$) to the magnetic field. In these cases, $\Lambda_{i3}=\Lambda_{3i}=0$ for both $i=1$ and $2$, so simple analytical expressions of the dispersion relations can be obtained.

\subsubsection{Perpendicular propagation}
When waves propagate perpendicular to the magnetic field, namely, when $k_\parallel=0$, the contributions by each charged species to the nonvanishing spatial components of the plasma response tensor are
\begin{eqnarray}
\nonumber
\hat{\Sigma}^{11}&=&-\frac{m\omega_p^2}{2m_0}\sum_{\varsigma=\pm 1}\kappa_{\varsigma}^2K_\varsigma^{(1)},\\
\nonumber
\hat{\Sigma}^{22}&=&\hat{\Sigma}^{11}-\frac{m\omega_p^2}{2m_0}\sum_{\varsigma=\pm 1}\kappa_{\perp}^2(K_\varsigma^{(0)}-2K_\varsigma^{(1)}),\\
\nonumber
\hat{\Sigma}^{12}&=&-\hat{\Sigma}^{21}=-\frac{m\omega_p^2}{2m_0}\sum_{\varsigma=\pm 1}(\pm i\varsigma)(\kappa_{\varsigma}^2-\kappa_{\perp}^2)K_\varsigma^{(1)},\\
\hat{\Sigma}^{33}&=&-\frac{m\omega_p^2}{m_0}.
\end{eqnarray}

The dispersion relations can be easily read out by substituting these into Eq. (\ref{eq:MagDisp}).  When the wave E-field is parallel to the background B-field, the wave is transverse. The dispersion relation of this ordinary electromagnetic wave (O-wave) is
\begin{equation}
\omega^2=\frac{m\omega_p^2}{m_0}+k_\perp^2.
\end{equation}
This is very similar to the dispersion relation of the O-wave in classical plasmas, except that the bare mass $m$ is now replaced by the ground state mass $m_0$. When the wave E-field is perpendicular to the background B-field, the longitudinal and transverse components of the wave are mixed by the off-diagonal components of the response tensor. Relativistic quantum cyclotron resonances (\ref{eq:Poles}) hybridize with the extraordinary electromagnetic wave (X-wave) by the dispersion relation
\begin{equation}\label{eq:Xwave}
(\omega^2+\hat{\Sigma}^{11})(\omega^2-k_\perp^2+\hat{\Sigma}^{22})-\hat{\Sigma}^{12}\hat{\Sigma}^{21}=0.
\end{equation}  
While the X-wave is more or less captured by classical plasma theories, cyclotron resonances, also known as the Bernstein waves, are absent in classical plasma theories when plasmas are cold \cite{Stix92}. In classical plasma theories, charged particles sample wave fields along their gyro-orbits. Bernstein resonances arise when their gyrofrequencies match the wave frequency. If plasma temperature is zero, cyclotron motion of classical particles stops and Bernstein resonances vanish consequently. However, this is not the case when quantum effects are taken into account. Using the uncertainty principle and the fact that the kinetic momentums $\mathbb{P}_{\mu}=-i\bar{D}_{\mu}$ do not commute $[\bar{D}_{\mu},\bar{D}_{\nu}]=-ie\bar{F}_{\mu\nu}$, 
it is easy to see that the gyromotion of quantum charged particles never stops. So Bernstein waves persist in a quantum plasma even when it is cold.

\begin{figure}[t]
	\renewcommand{\figurename}{FIG.}
	\includegraphics[angle=0,width=8.6cm]{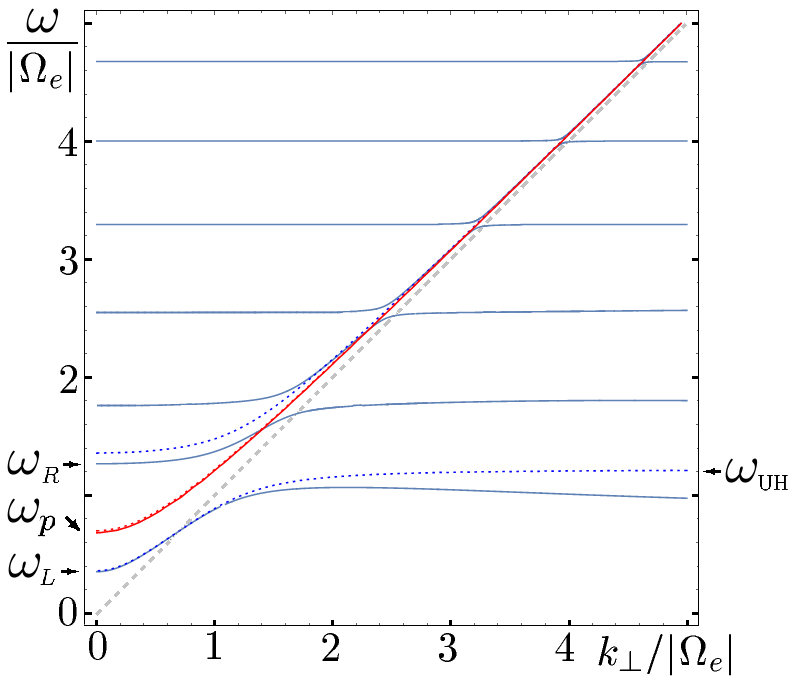}
	\caption{Perpendicular wave dispersion relations in a magnetized, cold, spinless electron gas with immobile ions as neutralizing background. Parameters used for making this plot are $\omega_{pe}/|\Omega_{e}|=0.7$ and $|\Omega_{e}|/m_e=0.1$. The solid curves are waves in a relativistic quantum plasma and the dashed curves are corresponding waves in a classical plasma. The red curves are the ordinary electromagnetic waves. The blue curves are the extraordinary electromagnetic waves hybridized with cyclotron resonances. The dashed gray line across the diagonal represents the light cone. Notice that near the light cone, wave dispersion relations in the relativistic quantum plasma asymptote to wave dispersion relations in the classical plasma. While the classical dispersion relations only capture the upper-hybrid resonance $\omega_{\text{UH}}$, the relativistic quantum dispersion relations capture all cyclotron resonances. Notice that gaps between Bernstein waves remain open even when the relativistic quantum plasma is cold. Also notice that cyclotron resonances are not harmonically spaced. The 5-th resonance occurs near $4\Omega$ instead of $5\Omega$ in this example.}
	\label{fig:PerpendicularDispersion}
\end{figure}

An example of wave dispersion relations for perpendicular propagation is plotted in Fig. \ref{fig:PerpendicularDispersion}. For the sake of clarity, we set the mass of ions to infinity, such that the immobile ions merely serve as a neutralizing background. By doing so, ion resonances and the gapless compressional Alfv$\acute{\text{e}}$n wave are removed, and what remain are the O-wave, the X-wave, and the relativistic quantum electron Bernstein waves. We only plot the low energy branches with $\omega,k_\parallel\ll m_e$, for which effects of the vacuum polarization can be safely ignored. Parameters used for making this plot are $\omega_{pe}/|\Omega_e|=0.7$ and $|\Omega_e|/m_e=0.1$. In Fig. \ref{fig:PerpendicularDispersion}, the solid curves are dispersion relations in a relativistic quantum plasma and the dashed curves are dispersion relations in a classical plasma. The solid and the dashed red curves almost overlap, since the dispersion relations of the relativistic quantum and the classical O-wave differ only in their mass gaps by the ratio $m/m_0\lesssim 1$. The blue curves are the X-waves hybridized with cyclotron resonances. While the classical dispersion relation only captures the upper-hybrid resonance at $\omega_{\text{UH}}=\sqrt{\Omega^2+\omega_p^2}$, the quantum dispersion relation captures all the cyclotron resonances, which are present even when the plasma is cold. Notice that cyclotron resonances are not harmonically spaced due to relativistic effects. As can be seen from Fig. \ref{fig:PerpendicularDispersion}, the 5-th resonance occurs near $4\Omega$ instead of $5\Omega$ in this example.  

The dispersion relations of relativistic quantum Bernstein waves may be approximated as follows. Denote $\omega_n:=\omega_{n,0}^{-}$, where $\omega_{n,k_\parallel}^{-}$ is defined in (\ref{eq:Poles}). Using property (\ref{eq:Kasympt}) of the K-function, the asymptotic behaviors of the plasma response tensor when $\omega\sim\omega_n$ are 
\begin{eqnarray}\label{eq:CyclotronRes}
\nonumber
\hat{\Sigma}^{11}&\sim&-\frac{m\omega_p^2}{m_0}\Big(1+\sigma_n\frac{\kappa_{+}^2}{\kappa_{+}^2-n}\Big),\\
\nonumber
\hat{\Sigma}^{22}&\sim&-\frac{m\omega_p^2}{m_0}\Big(1+\sigma_n\frac{\kappa_{+}^2-\kappa_\perp^2(2-\kappa_\perp^2/n)}{\kappa_{+}^2-n}\Big),\\
\hat{\Sigma}^{12}&=&-\hat{\Sigma}^{21}\sim-i\frac{m\omega_p^2}{m_0}\sigma_n\frac{\kappa_\perp^2-\kappa_{+}^2}{\kappa_{+}^2-n},
\end{eqnarray}
where $\sigma_{n+1}=(\kappa_\perp^2)^n\exp(-\kappa_\perp^2)/n!$. A good approximation of wave dispersion relations may be obtained by substituting the above expressions into (\ref{eq:Xwave}), keeping all the even powers of $\omega$ intact, and replacing odd powers $\omega^{2l+1}\rightarrow\omega^{2l}\omega_n$, such that the asymptotic behavior $\omega\sim k_\perp$ near the light cone is respected. To lowest order, the approximated dispersion relation near the resonance $\omega_n$ is
\begin{eqnarray}\label{eq:ApproxBernstein}
\omega^2&\sim&\frac{1}{2}\Big[\Big(\omega_n^2+\xi_n^2+k_\perp^2+\frac{m\omega_p^2}{m_0}\Big)\\
\nonumber
&&\pm\sqrt{\Big(\omega_n^2+\xi_n^2-k_\perp^2-\frac{m\omega_p^2}{m_0}\Big)^2+4\omega_n^2\xi_n^2}\hspace{4pt}\Big],
\end{eqnarray}
where $\xi_n^2=n|\Omega|m^2\omega_p^2\sigma_n/2\omega_nm_0^2$ is a function of $k_\perp$. The gaps between branches of relativistic quantum Bernstein waves are controlled by the factor $\omega_n\xi_n$. Notice that the gaps remain open even when plasmas are cold.

\begin{figure}[t]
	\renewcommand{\figurename}{FIG.}
	\includegraphics[angle=0,width=6.6cm]{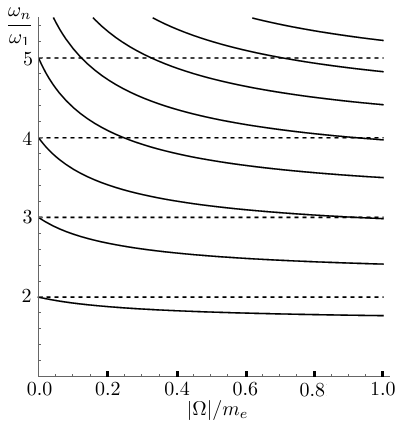}
	\caption{Ratios of the frequencies of cyclotron harmonics to the frequency of the fundamental. The solid curves are ratios when relativistic quantum effects are taken into account, and the dashed curves are the classical ratios. Quantum effects sustain cyclotron resonances even when the plasma is cold. Relativistic effects space resonances anharmonically even when the magnetic field is uniform. Notice that the anharmonicity is larger for stronger magnetic field and higher harmonic order.}
	\label{fig:HarmonicRatio}
\end{figure}

Cyclotron resonances in strongly magnetized plasmas are anharmonically spaced. In uniform magnetic fields, the anharmonicity is due to relativistic effect, which redshifts the cyclotron resonance $\omega_{n}$ from its classical value $n|\Omega|$. The redshift is significant when either the B-field is strong or the cyclotron order $n$ is large. More specifically, the redshift is comparable to the gyrofrequency, namely, $n|\Omega|-\omega_n\gtrsim|\Omega|$ when
\begin{equation}\label{eq:CycloRedShift}
n\gtrsim\sqrt{\frac{2m}{|\Omega|}}=\sqrt{\frac{2m^2c^2}{eB_0\hbar}}\approx 9.4\times\sqrt{\frac{10^{12}\hspace{3pt}\text{G}}{B_0}}.
\end{equation}
The ratio of the frequency of cyclotron harmonics to the frequency of the fundamental is plotted in Fig. \ref{fig:HarmonicRatio}. In the figure, the solid curves are ratios when relativistic quantum effects are taken into account, and the dashed lines are classical ratios. Using expression of $\omega_n^{-}$ in Eq. (\ref{eq:Poles}), it is easy to see when in the weak field limit $|\Omega|\ll m$, cyclotron resonances are harmonically spaced $\omega_n\sim n\omega_1$. While in the strong field limit $|\Omega|\gg m$, cyclotron resonances are anharmonically spaced with $\omega_n\sim \sqrt{n}\omega_1$. 

When magnetized plasmas are illuminated by external light sources, cyclotron resonances appear as absorption features. These anharmonic cyclotron absorption features have been observed in the spectra of a number of X-ray pulsars \cite{Heindl00,Santangelo99,Tsygankov06,Tsygankov07,Pottschmidt05,Makishima90}, whose polar caps serve as external light sources that illuminate their magnetospheres, which are made of magnetized plasmas.  The frequencies of line centers are determined by $\omega_{n,k_\parallel}^{-}$, which is given by Eq. (\ref{eq:Poles}). By fitting locations of absorption lines to Eq. (\ref{eq:Poles}), the line-averaged magnetic field can be determined. As has been noted in the previous paragraph, the spacing between absorption lines becomes anharmonic when Eq. (\ref{eq:CycloRedShift}) is satisfied. For X-ray pulsars, $B_0\sim10^{12}\hspace{3pt}\text{G}$, and the redshift starts to be significant from the 9-th resonance. The width and optical depth of the absorption lines are determined by gaps between branches of Bernstein waves, as well as the inhomogeneities of the plasma. Qualitatively, as can be seen from Fig. \ref{fig:PerpendicularDispersion}, lower Bernstein branches have larger gaps, resulting in wider absorption lines with larger optical depth. Quantitatively, when plasma density and magnetic field profile are known, the absorption line shapes can be found by solving the radiative transfer equations \cite{Meszaros92}, in which photons advection is governed by the dispersion relation (\ref{eq:MagDisp}). Conversely, when the absorption line shapes are measured, the plasma and magnetic field profile can be retrieved by solving the inverse problem. Thus a new era in astrophysics has been opened, in which the profile and evolution of the magnetosphere of an X-ray pulsar can be measured as it accretes materials from its companion star. 


\subsubsection{Parallel propagation}
When waves propagate parallel to the magnetic field, namely, when $k_\perp=0$, the nonvanishing spatial components of the plasma response tensor can be written as
\begin{eqnarray}
\nonumber
\hat{\Sigma}^{11}&=&\hat{\Sigma}^{22}=\omega^2(S-1),\\
\hat{\Sigma}^{12}&=&-\hat{\Sigma}^{21}=-i\omega^2D,\\
\nonumber
\hat{\Sigma}^{33}&=&\omega^2(P-1).
\end{eqnarray}
In the above expressions, $S=(R+L)/2$, $D=(R-L)/2$, and $P$ are notations of permittivities typically used in classical plasma physics. Using these notations, the dispersion relations of the right-handed circularly polarized electromagnetic R-wave, the left-handed circularly polarized electromagnetic L-wave, and the longitudinal electrostatic P-wave are
\begin{equation}
R=n_\parallel^2,\qquad L=n_\parallel^2, \qquad P=0.
\end{equation}
where $n_\parallel=k_\parallel/\omega$ is the refractive index. Writing summations over charged species explicitly, the permittivities 
\begin{eqnarray}
\nonumber
R&=&1-\sum_{s}\frac{m_s\omega_{ps}^2}{m_{s0}\omega^2}\frac{\omega^2-k_\parallel^2\mp 2m_{s0}\omega}{\omega^2-k_\parallel^2\mp 2(m_{s0}\omega+m_s\Omega_s)},\\
\nonumber
L&=&1-\sum_{s}\frac{m_s\omega_{ps}^2}{m_{s0}\omega^2}\frac{\omega^2-k_\parallel^2\pm 2m_{s0}\omega}{\omega^2-k_\parallel^2\pm 2(m_{s0}\omega-m_s\Omega_s)},\\
P&=&1-\sum_{s}\frac{m_s\omega_{ps}^2}{m_{s0}}\frac{\omega^2-k_\parallel^2-4m_{s0}^2}{(\omega^2-k_\parallel^2)^2-4m_{s0}^2\omega^2}.
\end{eqnarray} 
In the expressions of $R$ and $L$, the upper and lower sign of $\mp$ and $\pm$ correspond to $e_sB_0>0$ and $e_sB_0<0$, respectively. Since particle energy is not quantized in the direction parallel to the magnetic field, the low energy limit is the classical limit. In the classical limit $\omega,k_\parallel,|\Omega_e|\ll m_e$, it is clear that the above expressions asymptote to their classical values. Consequently, wave dispersion relations in relativistic quantum plasmas asymptote to those in classical plasmas.

\begin{figure}[t]
	\renewcommand{\figurename}{FIG.}
	\includegraphics[angle=0,width=8.6cm]{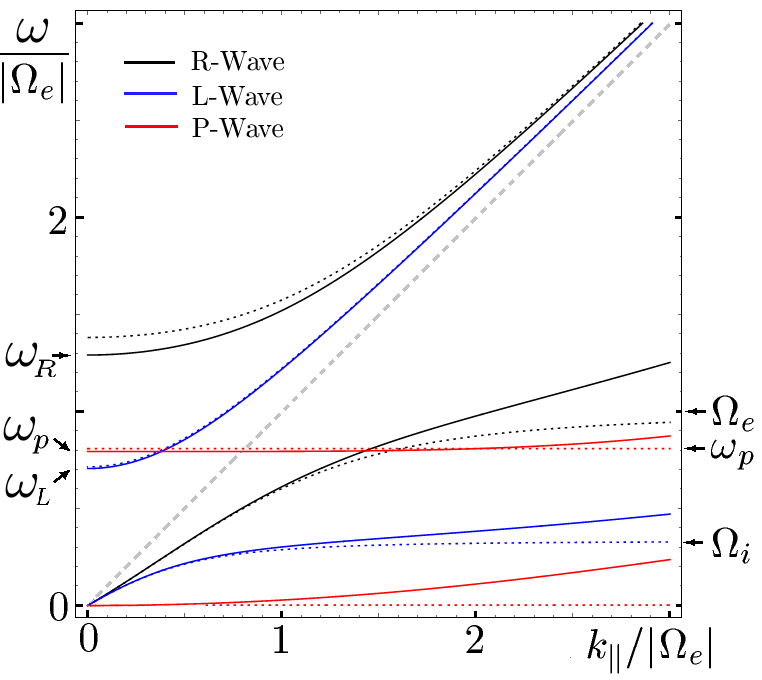}
	\caption{Parallel wave dispersion relations in a cold, magnetized, quasineutral, spinless electron-ion plasma. For various effects to be visible on the scale of this figure, parameters used for making this plot are $\omega_{pe}/|\Omega_{e}|=0.7$, $|\Omega_{e}|/m_e=0.1$ and $m_i/m_e=3$. The solid curves are waves in a relativistic quantum plasma and the dashed curves are corresponding waves in a classical plasma. The black and blue curves are the right- and left-handed circularly polarized electromagnetic waves, respectively. The red curves are the longitudinal electrostatic waves, which include a gapped Langmuir wave and a gapless acoustic wave. The dashed gray line across the diagonal represents the light cone. Notice that near the light cone, wave dispersion relations in the relativistic quantum plasma asymptote to wave dispersion relations in the classical plasma.}
	\label{fig:ParallelDispersion}
\end{figure}

An example of wave dispersion relations for parallel propagation in a quasineutral electron-ion plasma is plotted in Fig. \ref{fig:ParallelDispersion}. Only low energy branches with $\omega,k_\parallel\ll m_e$ are plotted, for which effects of the vacuum polarization can be safely ignored. In the figure, the solid curves are wave dispersion relations in a relativistic quantum plasma and the dashed curves are corresponding wave dispersion relations in a classical plasma. The black and blue curves are the right- and left-handed circularly polarized EM waves, respectively. The red curves are the longitudinal electrostatic waves, which include a gapped Langmuir wave and a gapless acoustic wave. For relativistic effects to be visible, the magnetic field is made strong such that $|\Omega_e|/m_e=0.1$. For ion effects to be visible, the mass of ions is chosen to be close to the electron mass with $m_i/m_e=3$. The ratio of the plasma frequency to the gyrofrequency is chosen to be $\omega_{pe}/|\Omega_e|=0.7$. It is easy to see that the relativistic quantum dispersion relations asymptote to the classical dispersion relations near the light cone. 

To get a sense of how large relativistic quantum corrections are, we can calculate the cutoff frequencies, the wave frequencies when the wave vector $\bm{k}=0$. In a single-species plasma, the approximated cutoff frequencies in the limit $\omega_p\sim|\Omega|\ll m$ are
\begin{eqnarray}\label{eq:MCutoff}
\nonumber
\frac{\omega_{R0}-\omega_R}{\omega_{R0}}&\sim&\frac{|\Omega|}{2m}\Big(1-\frac{\Omega}{\sqrt{\Omega^2+4\omega_p^2}}\Big),\\
\nonumber
\frac{\omega_{L0}-\omega_L}{\omega_{L0}}&\sim&\frac{|\Omega|}{2m}\Big(1+\frac{\Omega}{\sqrt{\Omega^2+4\omega_p^2}}\Big),\\
\frac{\omega_{P0}-\omega_P}{\omega_{P0}}&\sim&\frac{|\Omega|}{4m}.
\end{eqnarray} 
Here $\omega_{R0}$ is the cutoff frequency of the R-wave in a classical plasma and $\omega_{R}$ is the cutoff frequency of the R-wave in a relativistic quantum plasma. Similar notations are used for the L-wave and the P-wave. As expected, relativistic quantum effects are large when $|\Omega|\sim m$, namely, when the energy scale of the background magnetic field is comparable to the rest energy of charged particles. 

\begin{figure}[t]
	\renewcommand{\figurename}{FIG.}
	\includegraphics[angle=0,width=8.6cm]{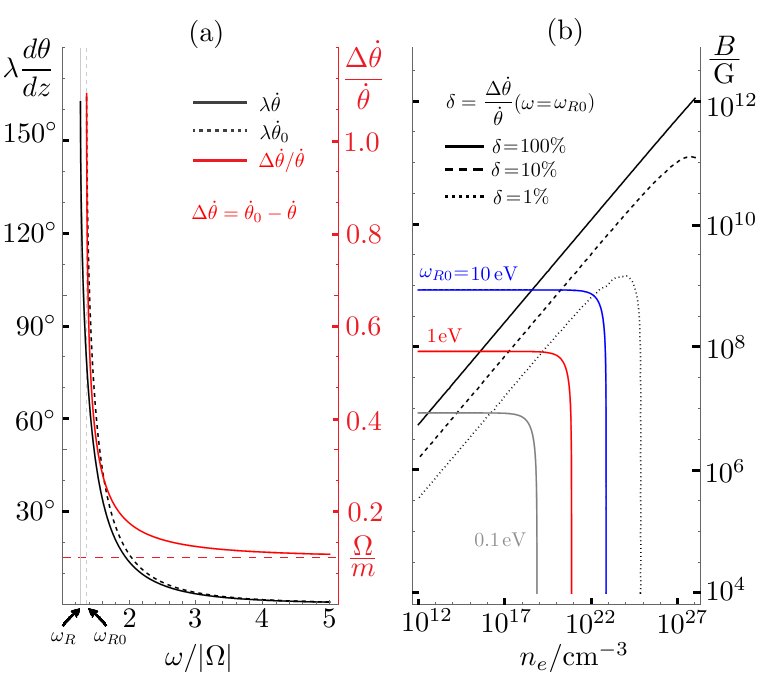}
	\caption{(a) Faraday rotation per vacuum wavelength in a cold, magnetized, spinless electron gas. The solid and dashed black curves are Faraday rotations in a relativistic quantum plasma and a classical plasma, respectively. The red curve is their relative difference. Parameters used for making this plot are $\omega_{pe}/|\Omega_{e}|=0.7$ and $|\Omega_{e}|/m_e=0.1$. Notice that near the classical cutoff $\omega_{R0}$, Faraday rotations in the relativistic quantum plasma and the classical plasma differ significantly. (b) Region in the $n_e-B$ space where relativistic quantum corrections are important. The regions above the solid, large-dashed and small-dashed black contours are regions where $\delta>100\%$, $ 10\%$ and $1\%$, respectively. The blue, red and gray contours are where $\omega_{R0}=10$ eV, $1$ eV and $0.1$ eV, respectively. These two sets of contours combined can be used to determine how important relativistic quantum corrections are in given conditions. See the main text for more details.}
	\label{fig:FaradayRotation}
\end{figure}

Although relativistic quantum corrections of cutoffs are small when $|\Omega|\ll m$, they can be important when magnified by singularities of refractive indices near cutoffs. For example, Faraday rotation, the rotation of the wave polarization axis caused by the phase velocity difference between the R-wave and the L-waves of the same frequency, can be substantially modified near cutoffs. Denote $\theta(z)$ the polarization angle of a linearly polarized wave as it propagates along the $z$-axis. Then Faraday rotation per vacuum wave length $\lambda=2\pi c/\omega$ is given by the well-known formula $\lambda\dot{\theta}=\pi(n_{\text{R}}-n_\text{L})$, where $n_\text{R}$ and $n_\text{L}$ are refractive indices of the R-wave and the L-wave, respectively. A comparison between Faraday rotations in a relativistic quantum plasma and a classical plasma is plotted in Fig. \ref{fig:FaradayRotation}(a), for parameters $\omega_{pe}/|\Omega_{e}|=0.7$ and $|\Omega_{e}|/m_e=0.1$. In the figure, the left axis is Faraday rotation per vacuum wavelength. The solid black curve is the Faraday rotation $\lambda\dot{\theta}$ in a relativistic quantum plasma and the dashed black curve is the Faraday rotation $\lambda\dot{\theta}_0$ in a classical plasma. The right axis of the figure is the relative difference $\dot{\theta}_0/\dot{\theta}-1$. As can be seen from the figure, while the relative difference asymptotes to a small number $|\Omega_e|/m_e$ when $\omega\gg|\Omega_e|$, it can be of order $1$ near the classical cutoff of the R-wave. Denote $\delta$ the relative difference at $\omega=\omega_{R0}$. The region in the $n_e-B$ space where $\delta$ is of order 1 is plotted in Fig. \ref{fig:FaradayRotation}(b). In the figure, the horizontal axis is the density of the electron gas and the vertical axis is the strength of the magnetic field. The region above the solid black contour is where $\delta>100\%$, the region above the large-dashed black contour is where $\delta>10\%$, and the region above the small-dashed black contour is where $\delta>1\%$. To facilitate reading of the figure, contours of $\omega_{R0}$ are also plotted. The blue contour is where $\omega_{R0}=10$ eV, the red contour is where $\omega_{R0}=1$ eV, and the gray contour is where $\omega_{R0}=0.1$ eV. The contours of $\delta$ and $\omega_{R0}$ combined can be used to determined how important relativistic quantum corrections are in a given situation. For example, the small-dashed black contour and the red contour intersect around $n_e\sim10^{19}\hspace{3pt}\text{cm}^{-3}$ and $B\sim10^{8}\hspace{3pt}\text{G}$. This means if laser with photon energy $\hbar\omega\sim 1$ eV is used to diagnose such plasma, then ignoring relativistic quantum effects will introduce $\sim1\%$ systematic error.

In laser plasma experiments, when lasers with frequencies close to classical cutoffs are used for diagnostics, relativistic quantum corrections of wave dispersion relations need to be taken into account in order to avoid systematic errors. As can be seen from Fig. \ref{fig:FaradayRotation}(a), if one tries to match data points on the relativistic quantum curve by shifting the classical curve, then $|\Omega_e|$ will have to be smaller than its true value, resulting in systematic errors. As the frequency of the diagnostic laser increases, the inferred magnetic field strength approaches its true value from the below. This is why the inferred magnetic field appears to increase with the frequency of the diagnostic laser when classical formulas are used. In experiments conducted by Tatarakis, Wagner and their coauthors \cite{Tatarakis02, Wagner04}, the magnetic field strength is determined from Cotton-Mouton effect, which depends on frequencies of cutoffs just as the Faraday rotation does. It is beyond the scope of this paper to go into details, but the peculiar dependence of the inferred magnetic field strength on the frequencies of diagnostic lasers can already be understood qualitatively as a consequence of relativistic quantum corrections of cutoff frequencies.

\section{\label{sec:conclusion}Conclusion and Discussion}
In this paper, we show that quantum field theory is an effective language for plasma physics. This language naturally incorporates relativistic and quantum effects without the necessity of manually adding patches to plasma models. To demonstrate the effectiveness of this language, we study wave propagation in scalar QED plasmas by calculating the 1-loop effective action of wave propagation. The 1-loop effective action contains all information of linear waves, and particularly, the dispersion relations. To order $e^2$, the effectively action (Eqs. \ref{1-loop}-\ref{vacPol}) captures effects of the polarization of the background plasma as well as the polarization of the vacuum. 

Using the 1-loop effective action and explicit expressions of cold plasma response tensors (Eqs. \ref{UMbk} and \ref{MagPol}), we show that all linear waves well-known in classical plasma theories can be recovered from relativistic quantum results when taking the classical limit. In unmagnetized plasmas (Fig. \ref{fig:UMDispersion}), the eigenmodes are the degenerate EM waves, the Langmuir wave, and the ion acoustic wave. When propagating perpendicular to the external B-field (Fig. \ref{fig:PerpendicularDispersion}), the eigenmodes are the electromagnetic O-wave and X-wave, and the electrostatic compressional Alfv$\acute{\text{e}}$n wave and the Bernstein waves. When propagating parallel to the external B-field (Fig. \ref{fig:ParallelDispersion}), the eigenmodes are the electromagnetic R-wave and L-wave, and the electrostatic Langmuir wave and acoustic wave. Since the Lagrangian of classical charged particles can be deduced from the relativistic quantum Lagrangian by making semiclassical approximations \cite{Ruiz15}, the agreement between relativistic quantum results and classical results in the classical regime is not coincidental. 

When relativistic quantum effects are important, we show that corrections to classical wave dispersion relations have profound observable consequences. In cold unmagnetized plasmas, relativistic quantum effects are important when either the plasma frequency $\hbar\omega_p$ or the wave vector $\hbar ck$ is comparable to electron rest energy $m_ec^2$. In these regimes, longitudinal waves propagate with nonzero group velocities even when the plasma is cold (Fig. \ref{fig:UMDispersion}). When propagating perpendicular the external B-field, quantum effects sustain cyclotron resonances even when the plasma is cold, and relativistic effects space these resonances anharmonically even when the magnetic field is uniform (Fig. \ref{fig:HarmonicRatio}). By providing explicit expressions of wave dispersion relations in strongly magnetized plasmas, we open up a new era in astrophysics where profiles of magnetospheres of X-ray pulsars can be retrieved from anharmonic cyclotron absorption features observed in their spectra. When propagating parallel to external B-fields, cutoff frequencies are modified differently by relativistic quantum effects. These modifications can be important even when relativistic quantum corrections are small, if lasers whose frequencies are close to the cutoffs are used for diagnostics. In this paper, we describe how Faraday rotation is modified (Fig. \ref{fig:FaradayRotation}a) and map out the parameter space where such modifications are important (Fig. \ref{fig:FaradayRotation}b). We show that for a large range of parameters relevant to laser plasma experiments, relativistic quantum corrections need to be taken into account in order to avoid systematic errors.       

Our descriptions of relativistic quantum plasmas are enabled by developing a quantum field theory with nontrivial background fields. The idea of separating classical backgrounds from quantum fluctuations is an extension to Furry's picture of strong field QED \cite{Furry51}. In addition to external EM fields, which are treated non-perturbatively in strong field QED, we also take into account of the existence of non-perturbative background charged particle fields. The formidable task of finding S-matrix elements by calculating quantum correlation functions whose end states contain infinitely many particles is reduced by incorporating effects of background charged particle fields directly into the Lagrangian. Such an incorporation, which has been attempted phenomenologically by Shvets \cite{Shvets95}, is made rigorous and systematic in this paper. We partition the fields into classical backgrounds and quantum fluctuations similar to what has been done by Raicher \textit{et al.} \cite{Raicher14}. We make further progress by simplifying the Lagrangian using the self-consistency of backgrounds, and developing the classical field theory to quantum level by including 1-loop effects. In this way, we thoroughly clarify the role of background fields and use boson plasmas as examples to demonstrate how nontrivial background fields can be treated in quantum field theory from first principle.

Using our formalism, further progress can be made by calculating higher order terms in the perturbation series. In the next order, effects such as particle collision and wave scattering will contribute. Without these effects, the linear response theory we have developed in this paper is a random phase approximation of relativistic quantum plasmas, in which particles respond collectively to the mean field instead of interacting directly with each other. To go beyond the random phase approximation, two steps are necessary. The first step is to solve the interacting many-body wave functions $\Phi_0$ and $\bar{A}$. The second step is to calculate the effective action to higher order in the perturbation series. For example, to describe wave propagation in thermal plasmas, one should first solve the thermal background states, which contain not only a thermal distribution of charged particles, but also a thermal distribution of wave quanta \cite{Bingham97}, and then calculate the propagator of the gauge boson to higher loops such that effects of collision between charged particles and scattering of waves are taken into account.

In the end, we want to point out the implication of this work for quantum field theory. An important consequence of including nontrivial background fields is a new mechanism for mass generation in gauge theories. In the Higgs mechanism \cite{Anderson63, Higgs64}, gauge bosons acquire masses through a gauge-symmetry-breaking mass term. For such a term to arise, the bare mass of the scalar field is required to be imaginary in order for the gauge symmetry to be spontaneously broken. In contrast, when nontrivial background fields are present, gauge bosons acquire masses through a gauge invariant response tensor (Eqs. \ref{bk} and \ref{bkConservation}). The response tensor endows gauge bosons with their masses, requiring that neither the bare mass of the scalar field be imaginary, nor the local gauge symmetry be broken. 

\begin{acknowledgments} 
The authors are grateful to I. Y. Dodin and D. E. Ruiz for valuable discussions. This research is supported by NNSA Grant No. DE-NA0002948 and DOE Research Grant No. DE-AC02-09CH11466.
\end{acknowledgments}

\begin{appendices}
\appendix
\renewcommand{\appendixname}{APPENDIX}

\section{\label{app:Green}\MakeUppercase{Green's Function in Symmetric Gauge}}
A spectral representation of the Green's function of charged scalar fields in a uniform magnetic field has been obtained in \cite{Melrose12}. In the Landau gauge 
\begin{equation}\label{eq:LandauGauge}
\bar{A}^{\mu}=(0,-B_0y,0,0),
\end{equation}
the Green's function can be written as
\begin{eqnarray}\label{eq:LandauGreen}
\nonumber
&G_{L}&(x,x')=\!\frac{\sqrt{2}}{r_0}\sum_{n=0}^{\infty}\int\frac{dq_0dq_\perp dq_\parallel}{(2\pi)^3}\\
&&\times\frac{ie^{i[q_0(t-t')-q_\perp(x-x')-q_\parallel(z-z')]}} {q_0^2-q_\parallel^2-m_n^2}\\
\nonumber
&&\times\psi_n\bm{\Big(}\frac{r_0}{\sqrt{2}}(q_\perp+eB_0y)\bm{\Big)}\psi_n\bm{\Big(}\frac{r_0}{\sqrt{2}}(q_\perp+eB_0y')\bm{\Big)}.
\end{eqnarray}
Here $G_L$ denotes the Green's function in the Landau gauge and $\psi_n(x)$ is the Hermite function. Using the completeness of the Hermite functions \cite{NIST10}
\begin{equation}
\sum_{n=0}^{\infty}\psi_n(x)\psi_n(y)=\delta(x-y),
\end{equation}
it is straightforward to check the Green's function (\ref{eq:LandauGreen}) solves the Schwinger$–-$Dyson equation (\ref{eq:SDGreen}) in the Landau gauge. This form of the Green's function is expanded by wave functions that are eigenfunctions of the $\mathbf{E}\times\mathbf{B}$ drift. These eigenfunctions are featured by free propagation in the $z$-direction, the direction of the $\mathbf{B}$ field; free propagation in the $x$-direction, the direction of $\mathbf{E}\times\mathbf{B}$ drift; and harmonic oscillation in the $y$-direction, the direction of the E-field. These features make the Green's function (\ref{eq:LandauGreen}) a convenient form for studying DC quantum Hall conductivity. But the loss of rotation symmetry in the perpendicular plane makes it inconvenient for studying AC wave phenomena.

To restore the rotation symmetry, we need to make a gauge transformation into the symmetric gauge (\ref{eq:SymmetricGauge}). The symmetric gauge is related to the Landau gauge (\ref{eq:LandauGauge}) by gauge transformation (\ref{BackgroundGauge}) with
\begin{equation}
\chi=-\frac{1}{2}B_0xy.
\end{equation} 
Under this gauge transformation, the Green's function is transformed by (\ref{eq:GaugeGreen}) as
\begin{equation}\label{eq:SLGreen}
G_{S}(x,x')=e^{-ieB_0(xy-x'y')/2}G_{L}(x,x'),
\end{equation}
where $G_S$ denotes the Green's function in the symmetric gauge. In the symmetric gauge, the eigenfunctions (\ref{eq:MagSingle}) are circular in the perpendicular plane, so the Green's function $G_S$, which can be expanded by these eigenfunctions, is also invariant under rotation around the $z$-axis.

To put $G_S$ in a manifestly rotational invariant form, we need the following relation between the Hermite functions and the Laguerre functions
\begin{equation}\label{eq:HermiteLaguerre}
\int dq e^{-2iqu}\psi_n(q+v)\psi_n(q-v)=\Lambda_n^{(0)}(2(u^2+v^2)).
\end{equation}
To prove this identity, recall the Hermite function 
\begin{equation}
\psi_n(x):=(2^nn!\sqrt{\pi})^{-1/2}e^{-x^2/2}H_n(x),
\end{equation}
where $H_n(x)$ is the Hermite polynomial. The Hermite polynomial satisfies
\begin{eqnarray}
&&H_n(x+y)=\sum_{k=0}^{n}\binom{n}{k}H_k(x)(2y)^{n-k},\\
&&\int dx H_n(x)H_m(x)e^{-x^2}=2^nn!\sqrt{\pi}\delta_{n,m}\hspace{2pt}.
\end{eqnarray}
Also recall that the Laguerre function
\begin{equation}
\Lambda_n^{(0)}(x):=L_n^{(0)}(x)e^{-x/2},
\end{equation}
where $L_n^{(0)}(x)$ is the Laguerre polynomial. The closed series of the Laguerre polynomial is
\begin{equation}
L_n^{(0)}(x)=\sum_{k=0}^{n}\binom{n}{k}\frac{(-x)^k}{k!}.
\end{equation}
Write $w=u+iv$ and $\bar{w}=u-iv$. With the change of variable $p=q+iu$, the LHS of the identity (\ref{eq:HermiteLaguerre}) becomes
\begin{eqnarray}
\nonumber
\text{LHS}&=&\frac{1}{2^nn!\sqrt{\pi}}\int dq e^{-q^2-v^2-2iqu}H_n(q+v)H_n(q-v)\\
\nonumber
&=&\frac{e^{-w\bar{w}}}{2^nn!\sqrt{\pi}}\int dp e^{-p^2}H_n(p-iw)H_n(p-i\bar{w})\\
\nonumber
&=&\frac{e^{-w\bar{w}}}{2^nn!}\sum_{k,l=0}^{n}\binom{n}{k}\binom{n}{l}(-2iw)^{n-k}(-2i\bar{w})^{n-l}2^kk!\delta_{k,l}\\
\nonumber
&=&e^{-w\bar{w}}\sum_{k=0}^{n}\binom{n}{k}\frac{(-2w\bar{w})^k}{k!}\\
\nonumber
&=&\text{RHS}.
\end{eqnarray}
We have thus proved the identity (\ref{eq:HermiteLaguerre}), which relates the integral of the Hermite functions to the Laguerre functions.

Using identity (\ref{eq:HermiteLaguerre}), we can carry out the $k_\perp$ integral in the Green's function (\ref{eq:LandauGreen}) and put the result in a rotational invariant form. With the following change of variables
\begin{eqnarray}
\nonumber
q&=&\frac{r_0}{\sqrt{2}}\Big(q_\perp+\frac{eB_0}{2}(y+y')\Big),\\
\nonumber
u&=&\frac{x-x'}{\sqrt{2}r_0},\\
\nonumber
v&=&\frac{r_0}{\sqrt{2}}\frac{eB_0}{2}(y-y'),
\end{eqnarray}
it is easy to see that the Green's function (\ref{eq:SLGreen}) can be put into its final form Eq. ({\ref{eq:MagGreen}}). 

Finally, using the completeness of the Laguerre functions \cite{NIST10} 
\begin{equation}
\sum_{n=0}^{\infty}\Lambda_{n}^{(0)}(x^2+y^2)=\pi\delta(x)\delta(y),
\end{equation}
it is straightforward to check that the Green's function (\ref{eq:MagGreen}) satisfies the Schwinger$–-$Dyson equation (\ref{eq:SDGreen}) in the symmetric gauge. We have thus found a convenient form of the Green's function for later calculations.

\section{\label{app:Derivative}\MakeUppercase{Covariant Derivatives in Symmetric Gauge}}
To evaluate the plasma response tensor ({\ref{bkPol}}) in the symmetric gauge (\ref{eq:SymmetricGauge}), we will need the following background gauge covariant derivatives 
\begin{eqnarray}
\nonumber
\bar{D}^{0}&=&\partial_t,\\
\nonumber
\bar{D}^{1}&=&-\partial_x+\frac{ieB_0}{2}y= -\cos\theta\frac{\partial}{\partial r}+\frac{\sin\theta}{r}\frac{\partial}{\partial\theta} \pm\frac{i\epsilon r}{r_0^2}\sin\theta,\\
\nonumber
\bar{D}^{2}&=&-\partial_y-\frac{ieB_0}{2}x= -\sin\theta\frac{\partial}{\partial r}-\frac{\cos\theta}{r}\frac{\partial}{\partial\theta}\mp\frac{i\epsilon r}{r_0^2}\cos\theta,\\
\bar{D}^{3}&=&-\partial_z.
\end{eqnarray}
Here we have used the identity $eB_0/2=\pm\epsilon/r_0^2$, where the upper and lower sign of $\pm$ correspond to $eB_0>0$ and $eB_0<0$, respectively. Since we denote $e$ the charge of the particle state, the charge of the antiparticle state is $-e$. This is why $\epsilon$ appears in the above expressions. Recall $\epsilon=+1$ for the particle state and $\epsilon=-1$ for the antiparticle state. 

To calculate the covariant derivatives of the wave functions, let us abbreviate the wave function as $\Psi=Me^{i\Theta}\eta^{l-n}\Lambda_n^{(l-n)}$, where $M$ is the constant amplitude, $\eta=r/r_0$ is the normalized radius, $\Theta=\epsilon[Et-pz\mp(l-n)\theta]$ is the phase, and the argument of $\Lambda_n^{(l-n)}$ is omitted. In terms of these abbreviations, the covariant derivatives of the wave function are
\begin{eqnarray}
\nonumber
\bar{D}^{0}\Psi&=&i\epsilon EMe^{i\Theta}\eta^{l-n}\Lambda_n^{(l-n)},\\
\nonumber
\bar{D}^{1}\Psi&=&\frac{M}{r_0}e^{i\Theta}[2\eta^{l-n+1}\Lambda_{n-1}^{(l-n+1)}\cos\theta\\
\nonumber
&&+e^{\pm i\epsilon\theta}\eta^{l-n-1}(\eta^2-l-n)\Lambda_{n}^{(l-n)}],\\
\nonumber
\bar{D}^{2}\Psi&=&\frac{M}{r_0}e^{i\Theta}[2\eta^{l-n+1}\Lambda_{n-1}^{(l-n+1)}\sin\theta\\
\nonumber
&&+e^{\pm i\epsilon(\theta-\pi/2)}\eta^{l-n-1}(\eta^2-l-n)\Lambda_{n}^{(l-n)}],\\
\bar{D}^{3}\Psi&=&i\epsilon pMe^{i\Theta}\eta^{l-n}\Lambda_n^{(l-n)}.
\end{eqnarray}
Here we have used the property of the Laguerre function $\Lambda_{n}^{'(\alpha)}(x)=-\Lambda_{n-1}^{(\alpha+1)}(x)-\Lambda_{n}^{(\alpha)}(x)/2$. Note $\Lambda_n^{(\alpha)}=0$ when $n<0$. In the above expressions, we see $\bar{D}^{2}\Psi$ can be obtained from $\bar{D}^{1}\Psi$ by replacing $\theta\rightarrow\theta-\pi/2$, which is expected from the rotation symmetry. 

To find the covariant derivatives of the Green's function, let us abbreviate $G=\Upsilon\mathcal{G}_ne^{iX}\Lambda_n^{(0)}$, where $\Upsilon=1/(\pi r_0^2)\sum_n\int d^2q/(2\pi)^2$ is the summation and integration prefactor, $\mathcal{G}_n(q_0,q_\parallel)=i/(q_0^2-q_\parallel^2-m_n^2)$ is the momentum space propagator in the $tz$-subspace, $X=\pm i\epsilon(xy'-x'y)/r_0^2+iq_0(t-t')-iq_\parallel(z-z')$ is the phase, and the argument of $\Lambda_n^{(0)}$ is omitted. Write $\eta_1=(x-x')/r_0$ and $\eta_2=(y-y')/r_0$, we find the covariant derivatives of the Green's function 
\begin{eqnarray}
\nonumber
\bar{D}^{0}G&=&\Upsilon\mathcal{G}_niq_0e^{iX}\Lambda_n^{(0)},\\
\nonumber
\bar{D}^{1}G&=&\Upsilon\mathcal{G}_n\frac{e^{iX}}{r_0}[2\eta_1\Lambda_{n-1}^{(1)}+(\eta_1\pm i\epsilon\eta_2)\Lambda_{n}^{(0)}],\\
\nonumber
\bar{D}^{2}G&=&\Upsilon\mathcal{G}_n\frac{e^{iX}}{r_0}[2\eta_2\Lambda_{n-1}^{(1)}+(\eta_2\mp i\epsilon\eta_1)\Lambda_{n}^{(0)}],\\
\bar{D}^{3}G&=&\Upsilon\mathcal{G}_niq_\parallel e^{iX}\Lambda_n^{(0)}.
\end{eqnarray} 
The covariant derivatives $\bar{D}^{'\mu}$ with respect to $x'$ can be found by direct calculations. Alternatively, recall $G'=-G^*$. We can also find $\bar{D}^{'\mu}$ derivatives using $\bar{D}^{'\mu}G=-(\bar{D}^{'*\mu}G')^*$. The above are all the background gauge covariant derivatives that are necessary for evaluating the plasma response tensor.

\section{\label{app:polarization}\MakeUppercase{Calculation of $\Pi_{2,\text{bk}}^{00}$}}
In this appendix, we calculate $\Pi_{2,\text{bk}}^{00}$ of a cold and uniformly magnetized plasma. For conciseness, we use abbreviations in Appendix \ref{app:Derivative}. We further abbreviate $\Psi_l=\Psi_{0,l,0}^{+}$, $\tau=t-t'$, $\zeta=z-z'$, $\bm{k}=(k_1,k_2)$, and $\bm{\eta}=(\eta_1,\eta_2)$. Substituting the density of states (\ref{eq:MagDOS}) into the effective single-boson wave function (\ref{eq:MagSingleEff}) and use it in place of the background field $\phi_0$ in the expression of the background polarization (\ref{bkPol}), we have
\begin{eqnarray}
\nonumber
\Pi_{2,\text{bk}}^{00}&=&e^2\sum_{l=0}^{\infty}\Upsilon(q_0+m_0)^2\Psi_l^*\Psi_l'\mathcal{G}_ne^{iX}\Lambda_n^{(0)}-c.c.\\
\nonumber
&=&\frac{e^2n_0}{2m_0}\Upsilon(q_0+m_0)^2\mathcal{G}_ne^{i[(q_0-m_0)\tau-q_\parallel\zeta]-\bm{\eta}^2/2}\Lambda_n^{(0)}-c.c.\\
&=&\frac{im\omega_p^2}{2m_0}\Upsilon e^{i(q_0\tau-q_\parallel\zeta)-\bm{\eta}^2/2}\Lambda_n^{(0)}\pi_n^{00},
\end{eqnarray}
where
\begin{eqnarray}
\nonumber
\pi_n^{00}(q_0,q_\parallel)&=&\frac{(q_0+2m_0)^2}{(q_0+m_0)^2-q_\parallel^2-m_n^2}\\
&+&\frac{(q_0-2m_0)^2}{(q_0-m_0)^2-q_\parallel^2-m_n^2}.
\end{eqnarray}
Note $\Pi_{2,\text{bk}}^{00}$ only depends on the difference between coordinates $r^{\mu}=(x-x')^{\mu}$. This is expected since the system is translational invariant.

In the Fourier space, since the system is translational invariant, we know $\hat{\Pi}(k,k')=(2\pi)^4\delta^{(4)}(k-k')\hat{\Pi}(k)$. Denoting $\bm{k}\bm{\eta}=k_1\eta_1+k_2\eta_2$, we have
\begin{eqnarray}
\nonumber
\hat{\Pi}_{2,\text{bk}}^{00}(k)&=&\int d^4r e^{ikr}\Pi_{2,\text{bk}}^{00}(r)\\
\nonumber
&=&\frac{im\omega_p^2}{2\pi m_0}\sum_{n=0}^{\infty}\int d\bm{\eta}e^{-ir_0\bm{k}\bm{\eta}-\bm{\eta}^2}L_n^{(0)}(\bm{\eta}^2)\pi_n^{00}(k_0,k_\parallel).
\end{eqnarray}
To calculate the integral, we need the following identities of the Laguerre and Hermite polynomials \cite{NIST10} 
\begin{eqnarray}
L_{n}^{(\alpha+\beta+1)}(x+y)&=&\sum_{l=0}^{n}L_{l}^{(\alpha)}(x)L_{n-l}^{(\beta)}(y),\\
L_{n}^{(-1/2)}(x^2)&=&\frac{(-1)^n}{2^{2n}n!}H_{2n}(x),
\end{eqnarray}
as well as the Fourier integral
\begin{equation}
\int dxe^{-ikx-x^2}H_{2n}(x)=\sqrt{\pi}(-1)^{n}k^{2n}e^{-k^2/4}.
\end{equation}
With these properties, we have
\begin{eqnarray}\label{eq:Fourier0}
\nonumber
I_n&:=&\int d\bm{\eta}e^{-i\bm{k}\bm{\eta}-\bm{\eta}^2}L_n^{(0)}(\bm{\eta}^2)\\
\nonumber
&=&\int d\bm{\eta}e^{-i\bm{k}\bm{\eta}-\bm{\eta}^2}\sum_{l=0}^{n}L_{l}^{(-1/2)}(\eta_1^2)L_{n-l}^{(-1/2)}(\eta_2^2)\\
\nonumber
&=&\frac{(-1)^n}{2^{2n}}\sum_{l=0}^{n}\frac{1}{l!(n-l)!}\int d\eta_1e^{-ik_1\eta_1-\eta_1^2}H_{2l}(\eta_1)\\
\nonumber
&&\times\int d\eta_2e^{-ik_2\eta_2-\eta_2^2}H_{2(n-l)}(\eta_2)\\
\nonumber
&=&\frac{\pi}{2^{2n}} e^{-\bm{k}^2/4}\sum_{l=0}^{n}\frac{k_1^{2l}k_2^{2(n-l)}}{l!(n-l)!}\\
&=&\frac{\pi}{n!} e^{-\bm{k}^2/4}\Big(\frac{\bm{k}^2}{4}\Big)^{n}.
\end{eqnarray}
Write $\bm{\kappa}^2=r_0^2\bm{k}^2/4$, then $\hat{\Pi}_{2,\text{bk}}^{00}(k)$ becomes
\begin{equation}\label{eq:sum0}
\hat{\Pi}_{2,\text{bk}}^{00}(k)=\frac{im\omega_p^2}{2m_0}e^{-\bm{\kappa}^2} \sum_{n=0}^{\infty}\frac{(\bm{\kappa}^2)^n}{n!}\pi_n^{00}(k_0,k_\parallel).
\end{equation}
Take the limit $B_0\rightarrow 0$, it is easy to see that the above expression recovers the polarization tensor $\hat{\Pi}_{2,\text{bk}}^{00}$ in the unmagnetized case. Notice that the pole of $\pi_n^{00}$ is weighted by $w_n(\bm{\kappa})=e^{-\bm{\kappa}^2}(\bm{\kappa}^2)^n/n!$, which is proportional to the strength of interaction between the plane wave with 4-momentum $k$ and particles in the $n$-th Landau level. The weighting factor maximizes at $\bm{\kappa}^2=n$. For large $n$, the maximum value scale as $w_n\sim1/\sqrt{2\pi n}$. We see waves couple more strongly to electrons in lower Landau levels.

The summation in (\ref{eq:sum0}) can be carried out using the confluent hypergeometric functions ${}_1F_1(a;b;z)$. Using the definition and properties of the confluent hypergeometric functions \cite{NIST10}, we have
\begin{eqnarray}\label{eq:sum1}
\nonumber
e^{-z}\sum_{n=0}^{\infty}\frac{z^n}{n!}\frac{1}{x+n}&=&\frac{e^{-z}}{x}{}_1F_1(x;x+1;z)\\
&=&\frac{1}{x}{}_1F_1(1;x+1;-z).
\end{eqnarray}
The $K$-function (\ref{eq:Kfunction}) is defined in such a way that
\begin{equation}\label{eq:Ksum}
e^{-z}\sum_{n=0}^{\infty}\frac{z^n}{n!}\frac{1}{x-n}=K(x,z).
\end{equation}
From this expression, it is easy to see that when $x\sim n$, where n is some integer, the $K$-function 
\begin{equation}\label{eq:Kasympt}
K(x,z)\sim \frac{z^n}{n!}\frac{e^{-z}}{x-n}.
\end{equation}
Using the $K$-function, the 00-component of the plasma polarization tensor can be written as
\begin{eqnarray}
\nonumber
\hat{\Pi}_{2,\text{bk}}^{00}(k)&=&\frac{im\omega_p^2}{2m_0}\frac{r_0^2}{4}[(k_0+2m_0)^2K(\kappa_{+}^2,\bm{\kappa}^2)\\
&&\hspace{30pt}+(k_0-2m_0)^2K(\kappa_{-}^2,\bm{\kappa}^2)],
\end{eqnarray}
where $\kappa_{\pm}^2=r_0^2(k_0^2-k_\parallel^2\pm 2k_0m_0)/4$. This is the final expression of $\hat{\Pi}_{2,\text{bk}}^{00}(k)$.

Other components of the plasma response tensor $\hat{\Pi}_{2,\text{bk}}^{\mu\nu}(k)$ can be calculated using similar methods. When calculating other components, one will encounter Fourier integrals similar to (\ref{eq:Fourier0}). The following property of the Laguerre polynomial is useful \cite{NIST10} 
\begin{equation}
L_{n}^{(\alpha+1)}(x)=\sum_{k=0}^{n}L_{k}^{(\alpha)}(x).
\end{equation}
Using this property, together with (\ref{eq:Fourier0}), all Fourier integrals that appear in the calculation of other components of $\hat{\Pi}_{2,\text{bk}}^{\mu\nu}$ can be calculated. For example,
\begin{eqnarray}
\nonumber
&&\int d\bm{\eta}e^{-i\bm{k}\bm{\eta}-\bm{\eta}^2}\eta_1L_{n-1}^{(1)}(\bm{\eta}^2)\\
&=&i\frac{\partial}{\partial k_1}\sum_{k=0}^{n-1}I_k=-\frac{i\pi }{2}e^{-\bm{\kappa}^2}k_1\frac{(\bm{\kappa}^2)^{n-1}}{(n-1)!}.
\end{eqnarray}

To carry out summations similar to (\ref{eq:sum1}), the following recurrence relation of the confluent hypergeometric function is useful \cite{NIST10} 
\begin{equation}\label{eq:confluent}
b\enspace{}_1F_1(a;b;z)=b\enspace{}_1F_1(a-1;b;z)+z\enspace{}_1F_1(a;b+1;z).
\end{equation} 
Using this recurrence relation, summations that appear in the calculation of other components of $\hat{\Pi}_{2,\text{bk}}^{\mu\nu}$ can be simplified. For example,
\begin{eqnarray}
\nonumber
e^{-z}\sum_{n=0}^{\infty}\frac{z^n}{n!}\frac{n}{x+n}&=&\frac{z}{x+1}{}_1F_1(1;x+2;-z)\\
&=&1-{}_1F_1(1;x+1;-z).
\end{eqnarray}

In terms of the $K$-function, the recurrence relation (\ref{eq:confluent}) becomes
\begin{equation}\label{eq:Kconfluent}
xK(x,y)-yK(x-1,y)=1.
\end{equation}
Using this identity, it is straightforward to show that the response tensor (\ref{MagPol}) satisfies the Ward$–-$Takahashi identity. 

\end{appendices}


\end{document}